\documentclass[a4paper,fleqn,usenatbib]{mnras}

\usepackage{newtxtext,newtxmath}
\usepackage[T1]{fontenc}
\usepackage{ae,aecompl}
\usepackage{graphicx}	% Including figure files
\usepackage{enumitem}
\usepackage{rotating}

\newcommand{\te}{\emph{TESS}}

\title[Photometry of evolved massive stars]{Red noise and pulsations in evolved massive stars}

\author[Y. Naz\'e et al.]{Ya\"el~Naz\'e\thanks{F.R.S.-FNRS Senior Research Associate, email: ynaze@uliege.be}, Gregor Rauw, and Eric Gosset\thanks{F.R.S.-FNRS Research Director}
\\
Groupe d'Astrophysique des Hautes Energies, STAR, Universit\'e de Li\`ege, Quartier Agora (B5c, Institut d'Astrophysique et de G\'eophysique), \\
All\'ee du 6 Ao\^ut 19c, B-4000 Sart Tilman, Li\`ege, Belgium
}

\pubyear{2020}

\begin{document}
\label{firstpage}
\pagerange{\pageref{firstpage}--\pageref{lastpage}}
\maketitle

\begin{abstract}
We examine high-cadence space photometry taken by the Transiting Exoplanet Survey Satellite (\te ) of a sample of evolved massive stars (26 Wolf-Rayet stars and 8 Luminous Blue Variables or candidate LBVs). To avoid confusion problems, only stars without bright {\it Gaia} neighbours and without evidence of bound companions are considered. This leads to a clean sample, whose variability properties should truly reflect the properties of the WR and LBV classes. Red noise is detected in all cases and its fitting reveals characteristics very similar to those found for OB-stars. Coherent variability is also detected for 20\% of the WR sample. Most detections occur at moderately high frequency (3--14\,d$^{-1}$), hence are most probably linked to pulsational activity. This work doubles the number of WRs known to exhibit high-frequency signals.
\end{abstract}


\section{Introduction}
Variability is an ubiquitous feature of stars, and massive objects are no exception in this respect. There is a large variety in the origin and amplitude of the changes. Some variations are intrinsic to the considered massive star. In this category, the most spectacular cases are the dramatic eruptions of Luminous Blue Variables (LBVs), which are massive stars in an unstable evolutionary stage. Less impressive but not less interesting are several other intrinsic processes. Rapid photospheric pulsations are ruled by the internal stellar structure \citep[e.g.][]{god17}. Slower rotational modulations arise from spots on the photosphere or from corotating features in the wind \citep[e.g.][]{Tahina}, as well as from a changing viewing angle on magnetically confined winds in oblique magnetic rotators \citep[e.g.][]{tow05}. Small-scale stochastic variations can also arise from clumps in the stellar winds \citep{mof88}. All those intrinsic variations are of utmost importance since they provide detailed information on the stellar properties and evolution. It is thus crucial to detect and characterize them.

With the advent of space facilities performing high-cadence photometry (Convection, Rotation et Transits plan\'etaires - {\it CoRoT}, Microvariability and Oscillations of STars - {\it MOST}, Bright-star Target Explorer - {\it BRITE}, Transiting Exoplanet Survey Satellite - \te), variability studies received a renewed interest. In the field of massive stars, this led to additional discoveries of isolated pulsations \citep{deg10,bri11}, frequency groups \citep[e.g.][]{bal20}, rotational modulation (e.g. $\zeta$\,Pup, \citealt{Tahina}), or red noise \citep[e.g.][]{blo11,rau19,bow20}. Most studies have focused on OB-stars, but a few more evolved objects have also been examined. For example, stochastic variability was detected for WR\,40 \citep{ram19}, WR\,103 \citep{mof08}, WR\,110 \citep{che11}, WR\,113 (\citealt{dav12} - see also a general summary by \citealt{len20}). Variations with rather long periods (days) were also found, either linked to eclipses (\citealt{sch19}, \citealt{dav12}) and ellipsoidal variability \citep{ric16} in binaries, or thought to be associated with rotationally-modulated corotating interacting regions in the winds (\citealt{che11}, \citealt{dav12}). Finally, a possibly stable periodicity near 2.45\,d$^{-1}$ was also reported in WR\,123 \citep{lef05} and near 0.7 and 1.3\,d$^{-1}$ for MWC\,314 \citep{ric16}, adding to a very small sample of high-frequency detections \citep[e.g.][]{ant95,ste97}.

In this paper, we continue these efforts by studying in depth the lightcurves of single evolved massive stars. To this aim, we first define a clean sample of single WR and LBV stars (Sect. 2), taking care to eliminate objects with bright visual neighbours and known bound companions. The periodograms of their lightcurves are then built, whose main features are examined in detail and compared to those of OB-stars (Sect. 3). The main results are finally summarized in Section 4.

\begin{table}
\centering
%  \scriptsize
\caption{List of targets by category, ordered by right ascension (R.A.). }
\label{listwr}
\setlength{\tabcolsep}{3.3pt}
\begin{tabular}{llcccc}
  \hline\hline
  Name & Sp.type & Sector & $\log(L_{\rm BOL}/L_{\odot})$ & $\log(\dot M)$ & $T_*$ \\
       &         &        &      &(M$_{\odot}$\,yr$^{-1}$) &(kK)\\
  \hline
\multicolumn{6}{l}{\it WR stars}\\
WR\,1	&WN4s	&17,18,24 &5.88	& --4.3	 &112.2\\
WR\,3	&WN3hw	&18$^*$   &5.56	& --5.4	 &89.1 \\
WR\,4	&WC5	&18$^*$   &5.71	& --4.37 &79   \\
WR\,5	&WC6	&18       &5.53	& --4.59 &79   \\
WR\,7	&WN4s	&7        &5.36	& --4.8	 &112.2\\
WR\,15	&WC6	&8,9      &5.99	& --4.14 &79   \\
WR\,16	&WN8h	&9,10     &5.72	& --4.6  &44.7 \\
WR\,17	&WC5	&9,10     &5.74	& --4.4	 &79   \\
WR\,23	&WC6	&10       &5.61	& --4.49 &79   \\
WR\,24	&WN6ha	&10$^*$,11&6.47	& --4.3	 &50.1 \\
WR\,40	&WN8h	&10$^*$,11&5.91	& --4.2	 &44.7 \\
WR\,52	&WC4	&11       &5.07	& --4.75 &112  \\
WR\,57	&WC8	&11,12    &5.75	& --4.5	 &63   \\
WR\,66	&WN8(h)	&12       &6.15	& --3.9	 &44.7 \\
WR\,78	&WN7h	&12       &5.8	& --4.5  &50.1 \\
WR\,79b	&WN9ha	&12$^*$   &5.825& --4.565&28.5 \\
WR\,81	&WC9	&12       &5.26	& --4.62 &45   \\
WR\,84	&WN7	&12       &5.36	& --4.8  &50.1 \\
WR\,92	&WC9	&12       &4.95	& --5	 &45   \\
WR\,96	&WC9d	&12       &  	&	 &     \\
WR\,130	&WN8(h)	&14       &6.25	& --4.2	 &44.7 \\
WR\,134	&WN6s	&14,15    &5.61	& --4.4	 &63.1 \\
WR\,135	&WC8	&14,15    &5.4	& --4.73 &63   \\
WR\,136	&WN6(h)	&14,15    &5.78	& --4.2	 &70.8 \\
WR\,138a&WN8-9h	&14,15    &5.3	& --4.7	 &40   \\
WR\,154	&WC6	&16,17    &5.91	& --4.26 &79   \\
  \hline
\multicolumn{5}{l}{\it LBVs} & $T_{\rm eff}$\\
\multicolumn{2}{l}{AG\,Car}           &10,11       & 6.14--6.22 & & 13--29\\
\multicolumn{2}{l}{WRAY\,15-751}      &10,11       & 5.91 & & 30.2 \\  
\multicolumn{2}{l}{P\,Cyg}            &14,15       & 5.70 & & 18.2 \\
\multicolumn{6}{l}{\it LBV candidates}\\
\multicolumn{2}{l}{HD\,80077}         &8$^*$,9$^*$ & 6.30 & & 17.0 \\
\multicolumn{2}{l}{\scriptsize 2MASS\,J16493770-4535592}&12   &       & &      \\
\multicolumn{2}{l}{$\zeta^1$\,Sco}    &12$^*$      & 6.10 & & 18.2 \\
\multicolumn{2}{l}{GRS\,G079.29+00.46}&14,15       & 6.30 & & 25.1 \\
\multicolumn{2}{l}{Schulte\,12}       &14,15       & 6.42 & & 12.9 \\
  \hline      
\end{tabular}

{\scriptsize $^*$ indicates 2\,min cadence data; physical properties come from \citet[and references therein]{naz12} for LBVs and candidates, and from the fit results of \citet{san19} for WC stars, \citet{boh99} for WR\,79b, \citet{gva09} for WR\,138a, and \citet{ham19} for other WN stars. The temperature $T_*$ of Wolf-Rayet stars is defined as the effective temperature related to the stellar luminosity and the stellar radius for an optical depth $\tau_{\rm Ross}=20$ via the Stefan-Boltzmann law \citep{san19,ham19}. } 
\end{table}

\begin{figure}
  \begin{center}
    \includegraphics[width=8cm]{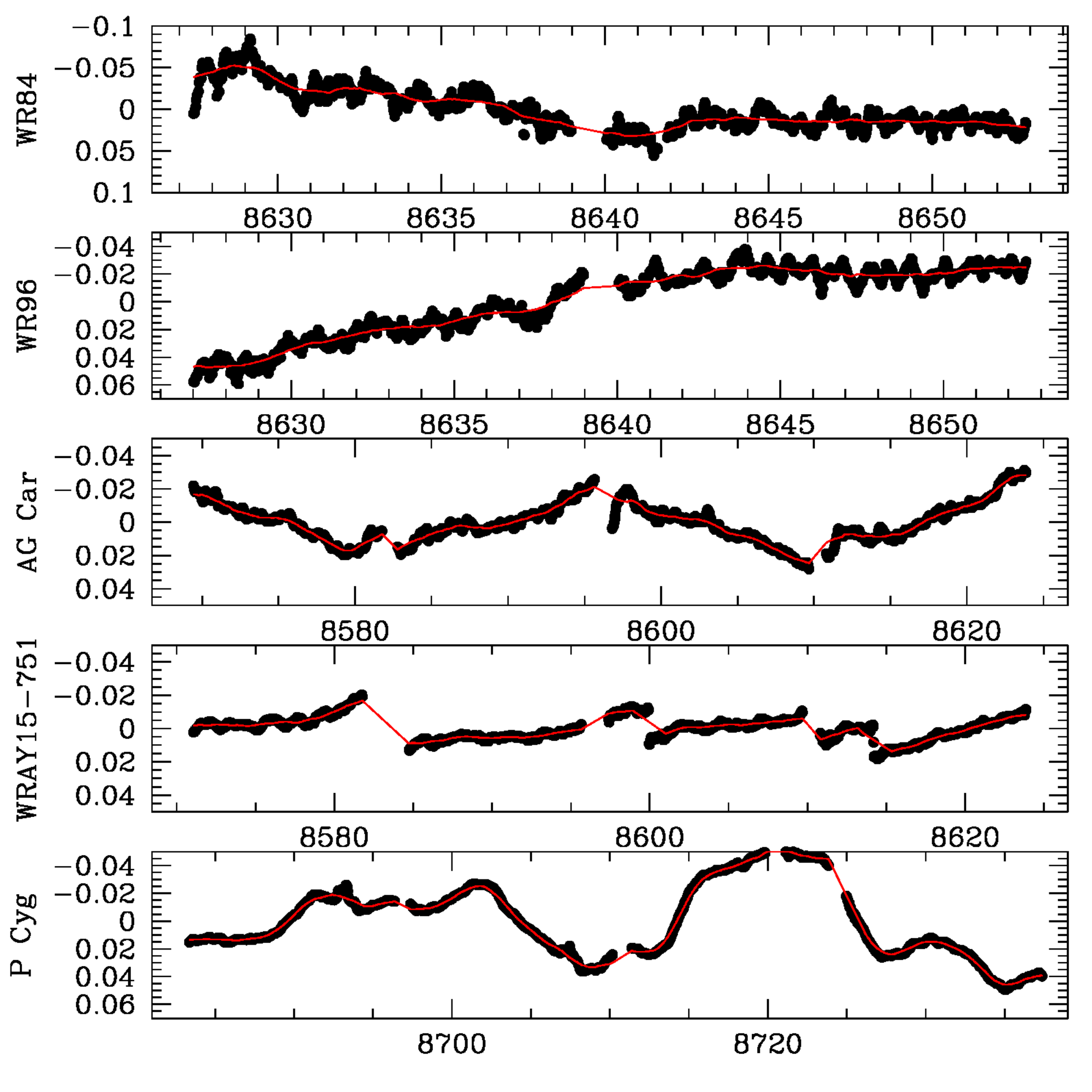}
    \includegraphics[width=8cm]{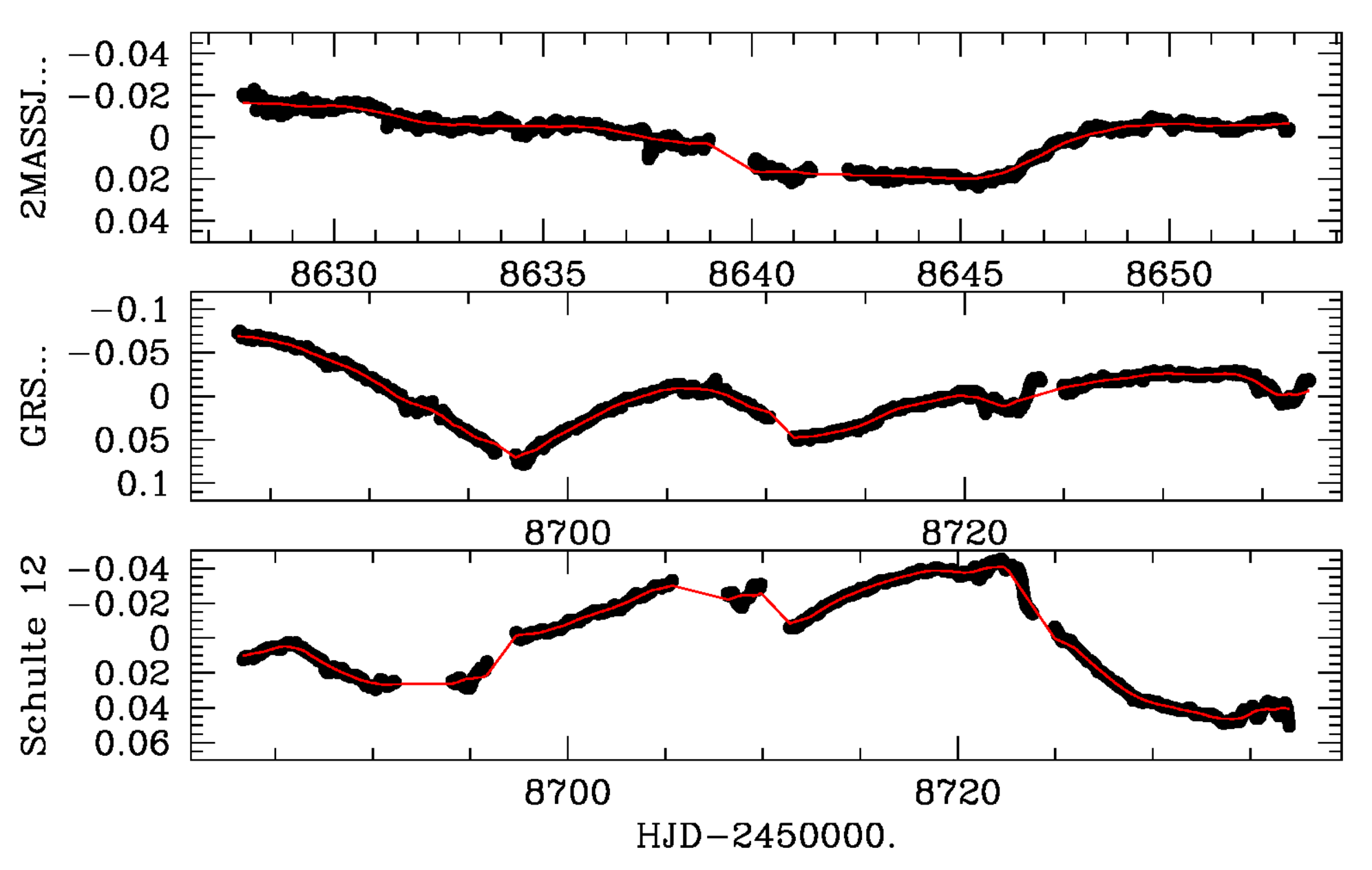}
  \end{center}
  \caption{Original lightcurves (in mag) and their long-term trends (in red) which were taken out for subsequent analysis.}
\label{detr}
\end{figure}

\section{The stars and their data}
\subsection{The sample}
To select the WR stars of our sample, we cross-correlated {\it Gaia}-DR2 \citep{gaia16,gaia18} with the WR catalog of \citet{vdh01} and its online extension\footnote{http://pacrowther.staff.shef.ac.uk/WRcat/} by P.A. Crowther. To avoid crowding issues, since \te\ has a 50\% ensquared-energy half-width of 21\arcsec, corresponding to one detector pixel, and since the photometry is extracted over several pixels, we discarded objects having bright ($\Delta G<2.5\,mag$) and close (within 1\arcmin ) neighbours. Of the remaining 76 WR stars, only 61 had available \te\ photometry. Since we are interested in the intrinsic characteristics of WR stars, we need to avoid contamination by bound companions too. Therefore, we further discarded all objects known to be spectroscopic binaries (SB1 or SB2) as well as those showing some evidence of multiplicity (presence of non-thermal radio emission, dust making, detected radial velocity shifts). This left 26 WR stars, nearly equally split between WC and WN types, assumed to be single (Table \ref{listwr}). For all targets but one (WR\,96), the physical parameters are known since they were derived using atmosphere modelling by \citet{san19} for WC stars, \citet{boh99} for WR\,79b, \citet{gva09} for WR\,138a, and \citet{ham19} for other WN stars. 

\begin{figure*}
  \begin{center}
\includegraphics[width=5.8cm]{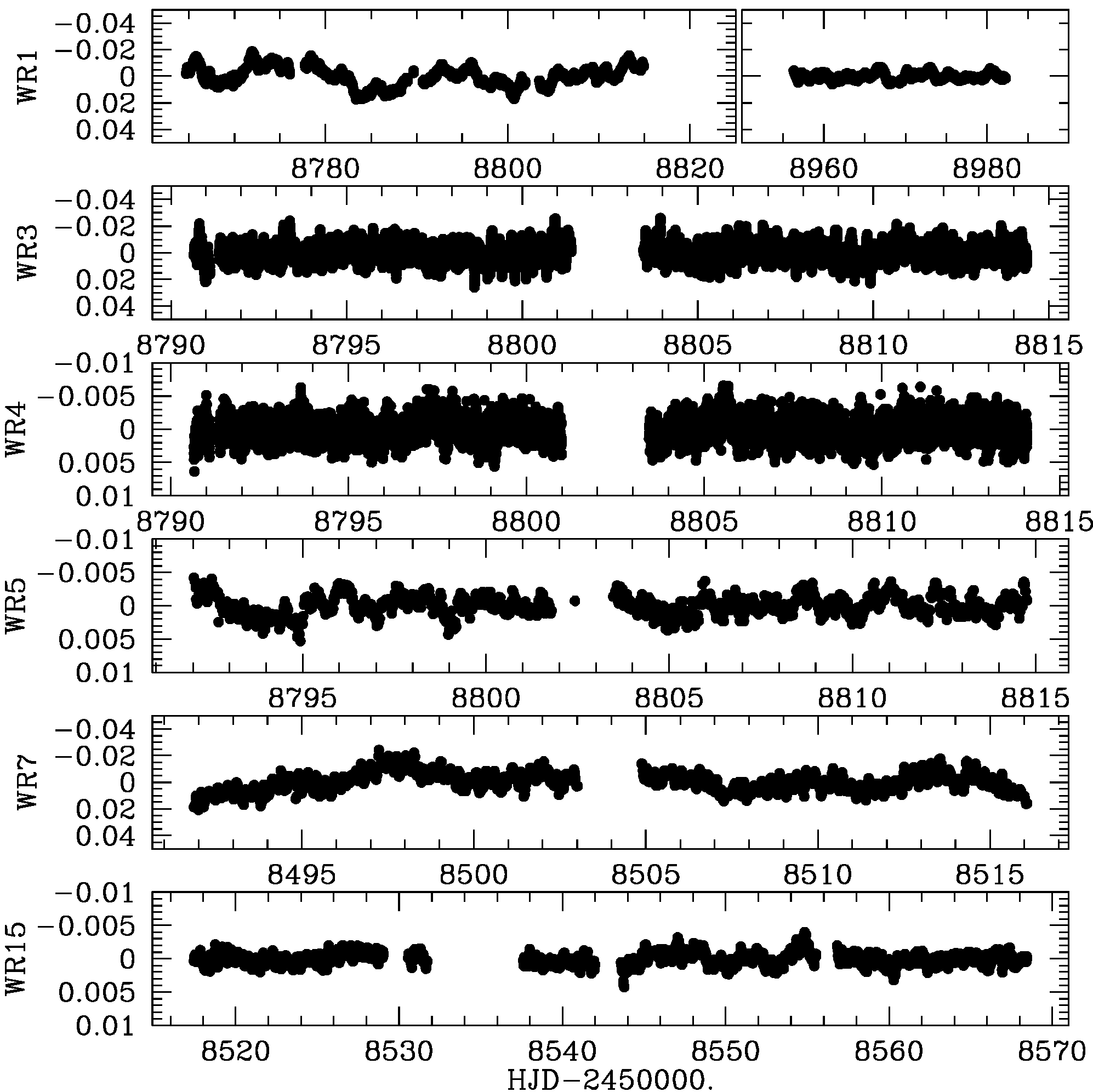}
\includegraphics[width=5.8cm]{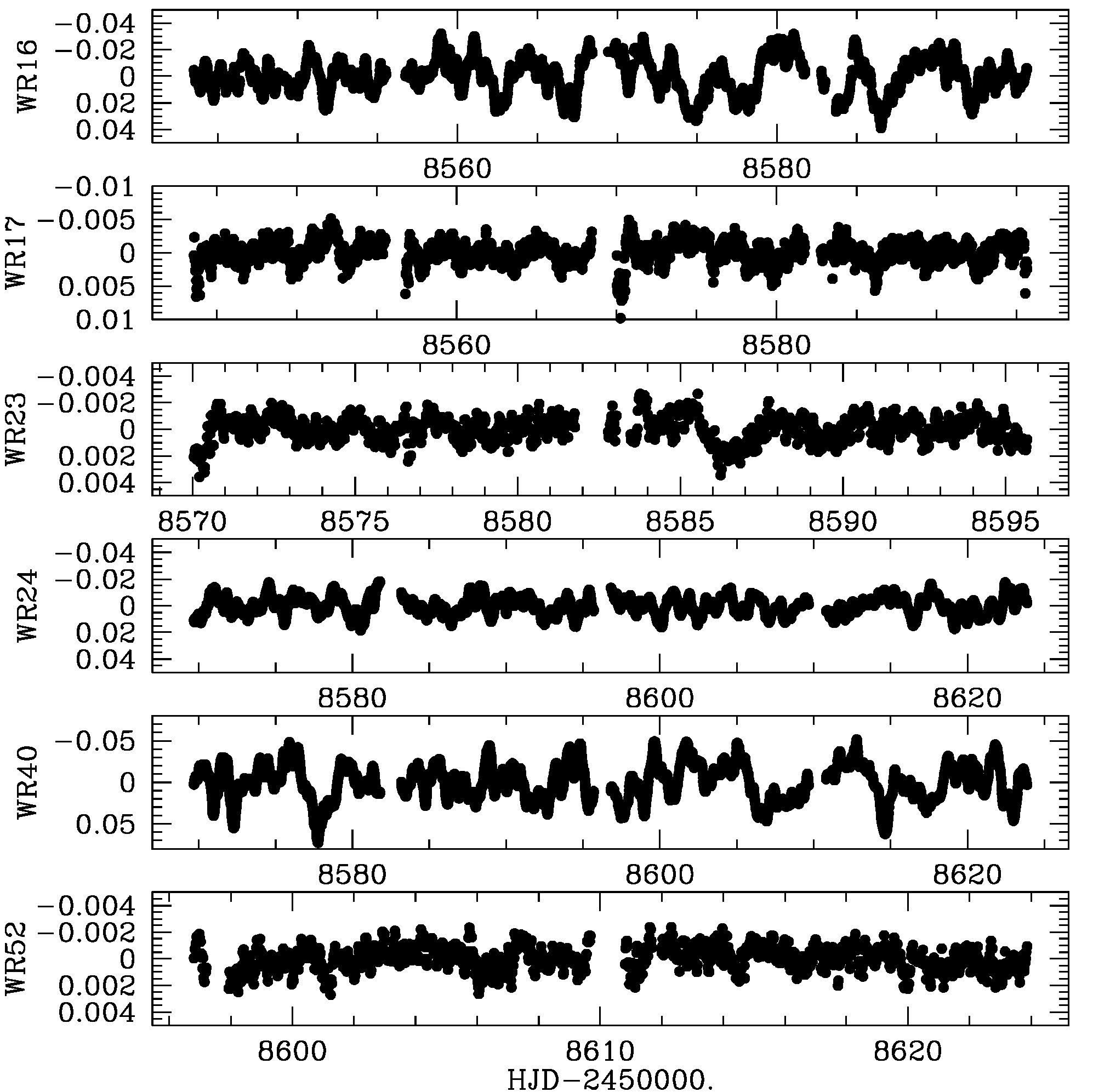}
\includegraphics[width=5.8cm]{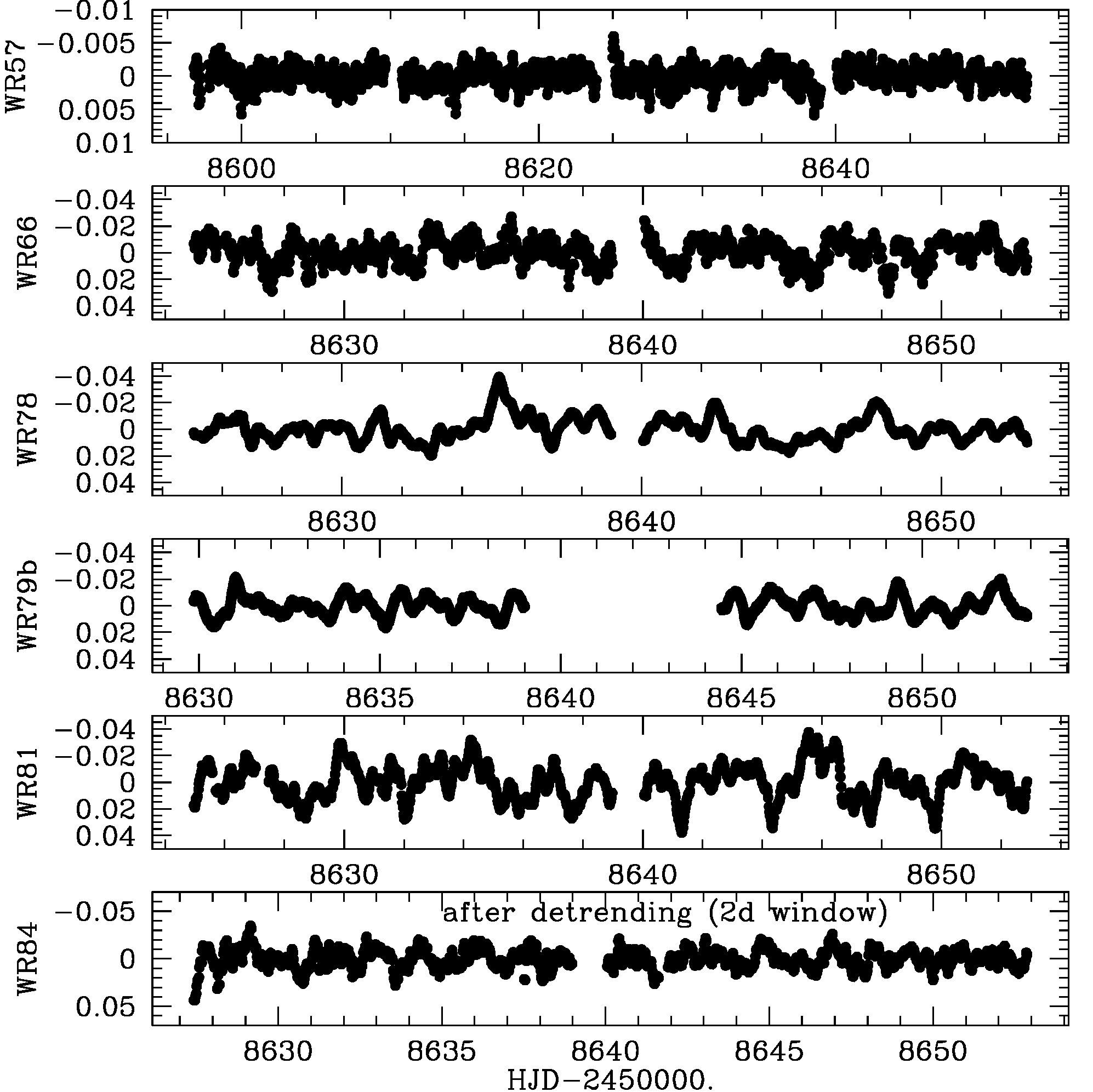}
\includegraphics[width=5.8cm]{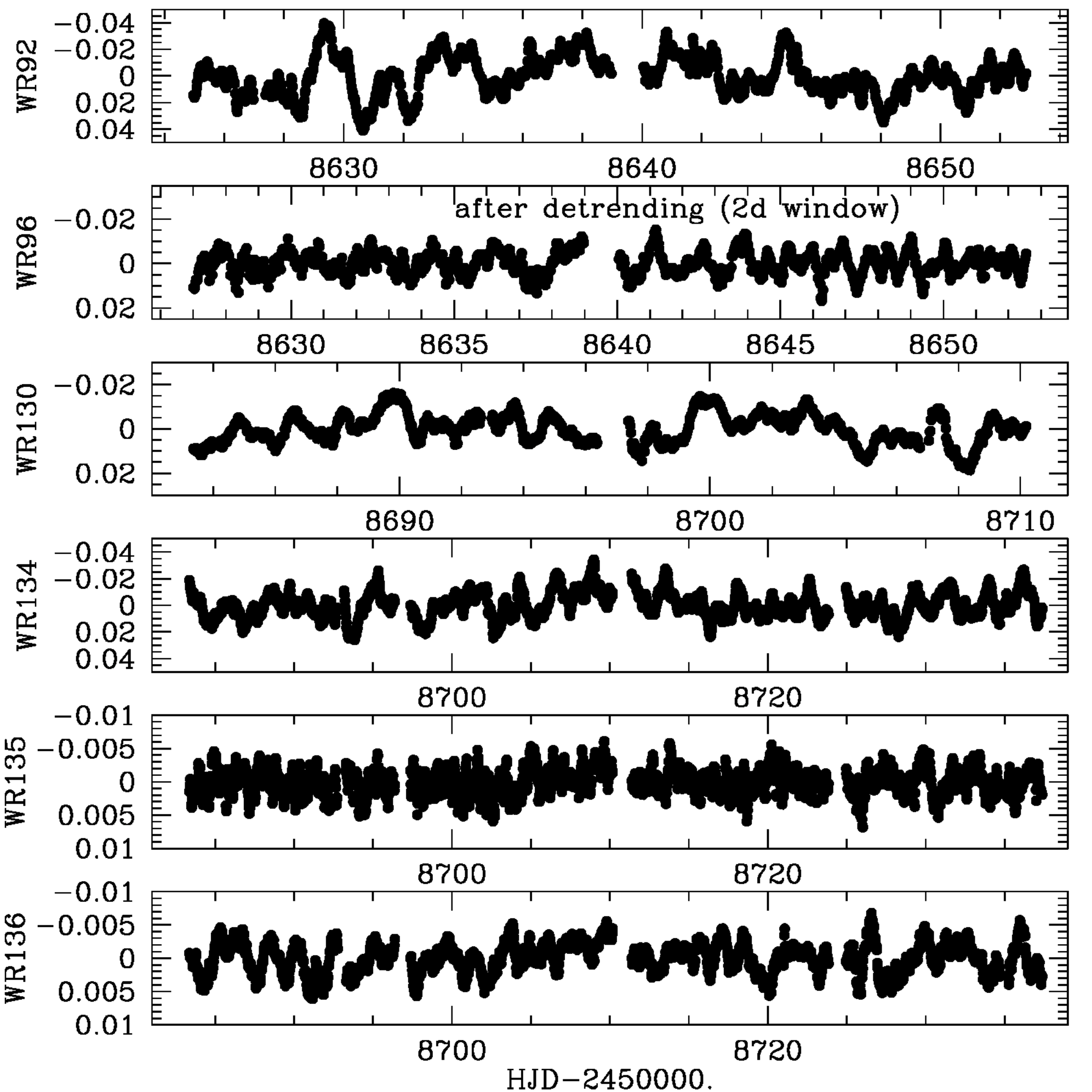}
\includegraphics[width=5.8cm]{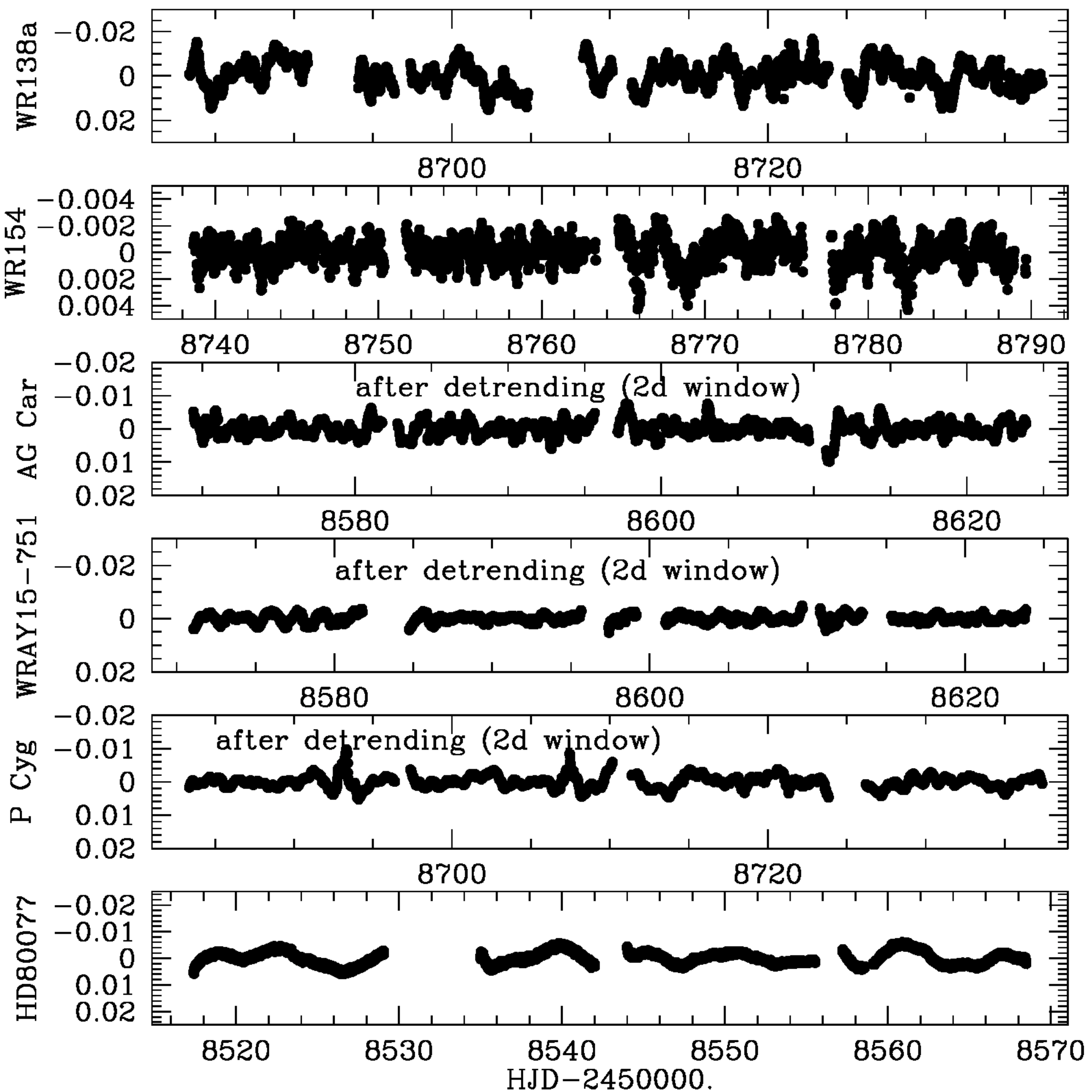}
\includegraphics[width=5.8cm]{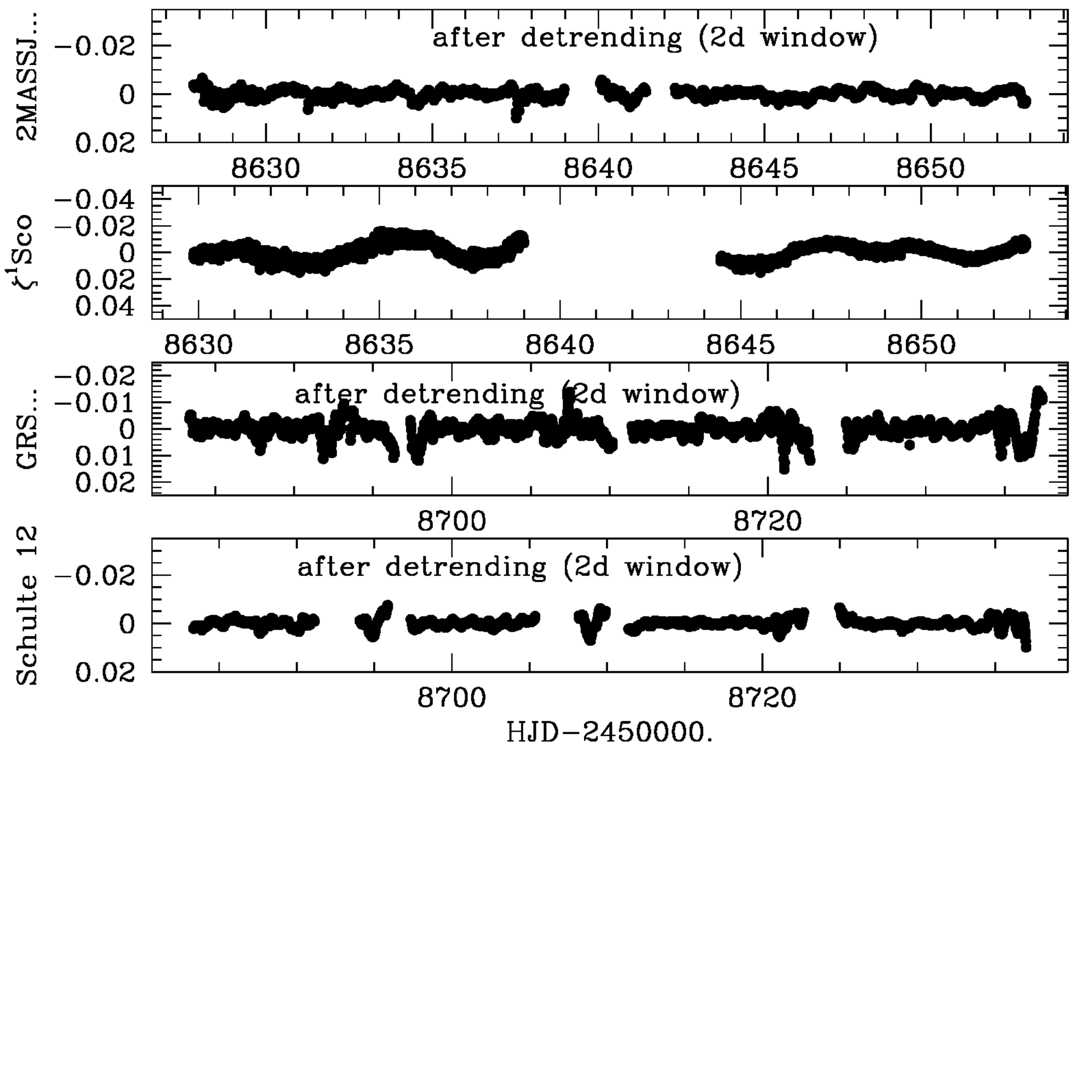}
  \end{center}
  \caption{\te\ lightcurves of the targets (in mag).}
\label{lcwr}
\end{figure*}

From the list of LBVs and LBV candidates (cLBVs) from \citet{naz12}, we discarded objects having bright ($\Delta G<2.5\,mag$) and close (within 1\arcmin ) {\it Gaia}-DR2 neighbours, as done for WR stars. Amongst the remaining stars, ten (3 LBVs and 7 candidates) were observed by \te. The multiplicity of LBVs is much less known than for WRs but amongst our targets, two certainly are binaries because they present periodic radial velocity variations and photometric changes: MWC\,314 \citep{ric16} and HD\,326823 \citep{ric11}. Having several components in a target can lead to confusion on the origin of the short-term photometric variability. Indeed, the pulsations detected by {\it MOST} for MWC\,314 are attributed to its stripped He-star companion \citep{ric16}. Therefore, we discard those two stars from our target list. Table \ref{listwr} provides the final selection, along with their stellar properties taken from \citet[and references therein]{naz12}. 

Note that the presence of very close neighbours is not yet mentioned in {\it Gaia}-DR2. However, the {\it Gaia}-DR2 available at ESA archives\footnote{https://gea.esac.esa.int/archive/} provides a ``Renormalised Unit Weight Error'' (RUWE, \citealt{lin16}) which should be close to unity if a good fitting of the astrometric observations was achieved by the single star model. For our sample, RUWE is close to one for all stars but WR\,66 (RUWE=14), WR\,79b (RUWE=3.7), and WR\,130 (RUWE=2.7). Furthermore, of the stars listed in Table \ref{listwr}, only one (WR\,66) possesses a neighbouring component in the {\it Hipparcos} catalogue: it is located 0.4\arcsec\ away from the WR star and is one magnitude fainter than the latter. Since it passes the chosen criteria, this star is kept in our target list but we remind that some caution should be applied for its results.

\subsection{The \te\ lightcurves}
Launched in April 2018, the \te\ satellite \citep{ric15} provides photometric measurements for $\sim$85\% of the sky. While sky images are taken every 30\,min, subarrays on preselected stars are read every 2\,min. Observations are available for at least one sector, corresponding to a duration of $\sim$25\,d. The main steps of data reduction (pixel-level calibration, background subtraction, flatfielding, and bias subtraction) are done by a pipeline similar to that designed for the {\it Kepler} mission.

For 2\,min cadence data, time-series corrected for crowding, the limited size of the aperture, and instrumental systematics are available from the MAST archives\footnote{https://mast.stsci.edu/portal/Mashup/Clients/Mast/Portal.html}. We kept only the best quality (quality flag=0) data. In our sample, only seven stars have such very high cadence data: WR\,3, 4, 24 (only sector 10), 40 (only sector 10), 79b, $\zeta^1$\,Sco and HD\,80077. 

For the other stars, individual lightcurves were extracted for each target from \te\ full frame images with 30\,min cadence. Aperture photometry was done on image cutouts of 50$\times$50 pixels using the Python package Lightkurve\footnote{https://docs.lightkurve.org/}. A source mask was defined from pixels above a given flux threshold (generally 10 Median Absolute Deviation over the median flux, but it was decreased for faint sources or increased if neighbours existed). The background mask was defined by pixels with fluxes below the median flux (i.e. below the null threshold), thereby avoiding nearby field sources. A principal component analysis (PCA) then helped correcting the source curve for the background contamination (including scattered light). The number of PCA components was set to five, except for WR\,40 and the (c)LBVs where a value of 2 provided better results. All data points with errors larger than the mean of the errors plus three times their 1$\sigma$ dispersion were discarded. In addition, a few isolated outliers and a few short temporal windows with sudden high scatter were also discarded. For example, for WR\,84 and 96, the first hundred frames ($HJD<2\,458\,627$) were affected by a very local and intense patch of scattered light, hence they were discarded. 

\begin{figure*}
  \begin{center}
\includegraphics[width=5.8cm]{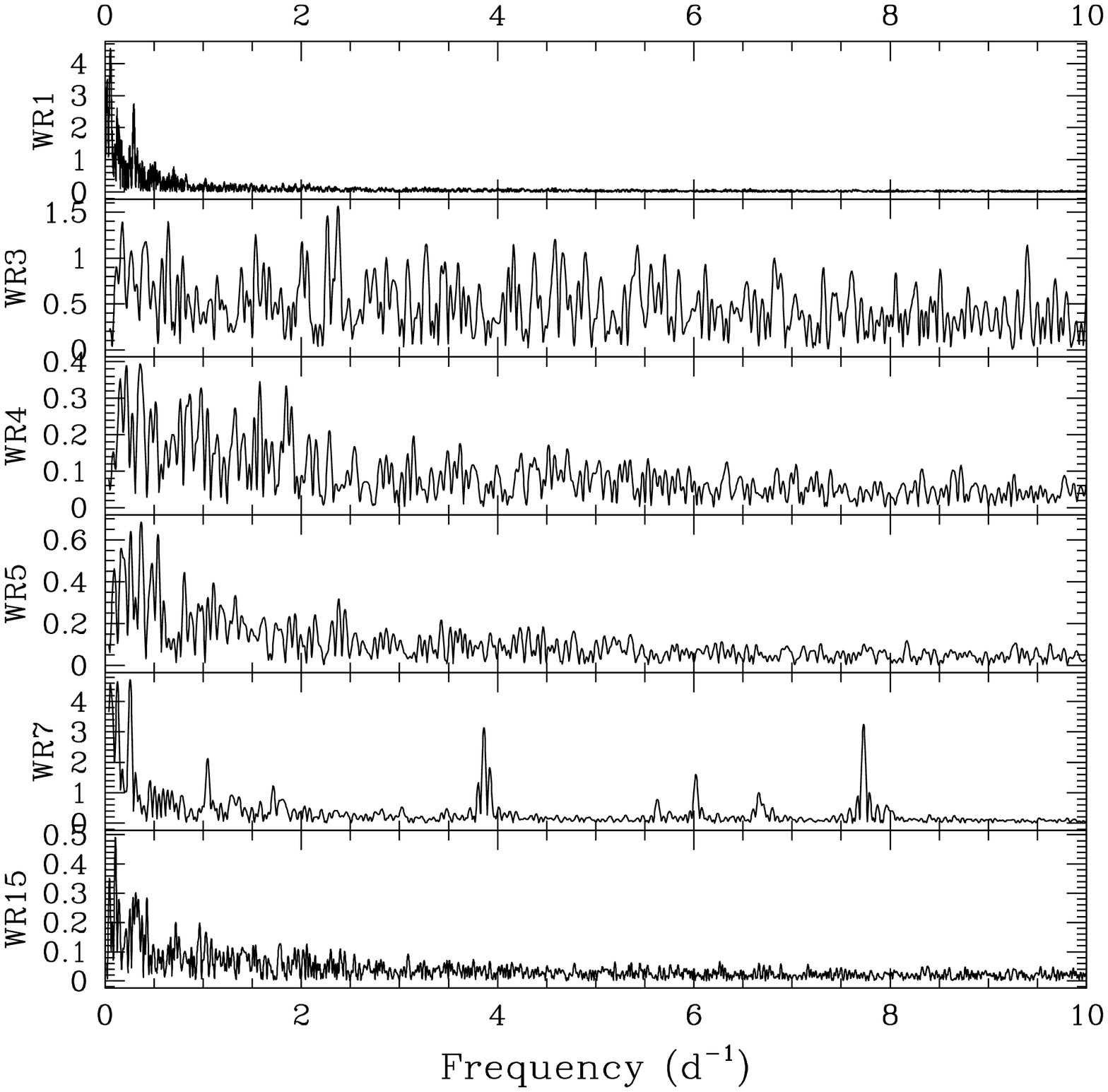}
\includegraphics[width=5.8cm]{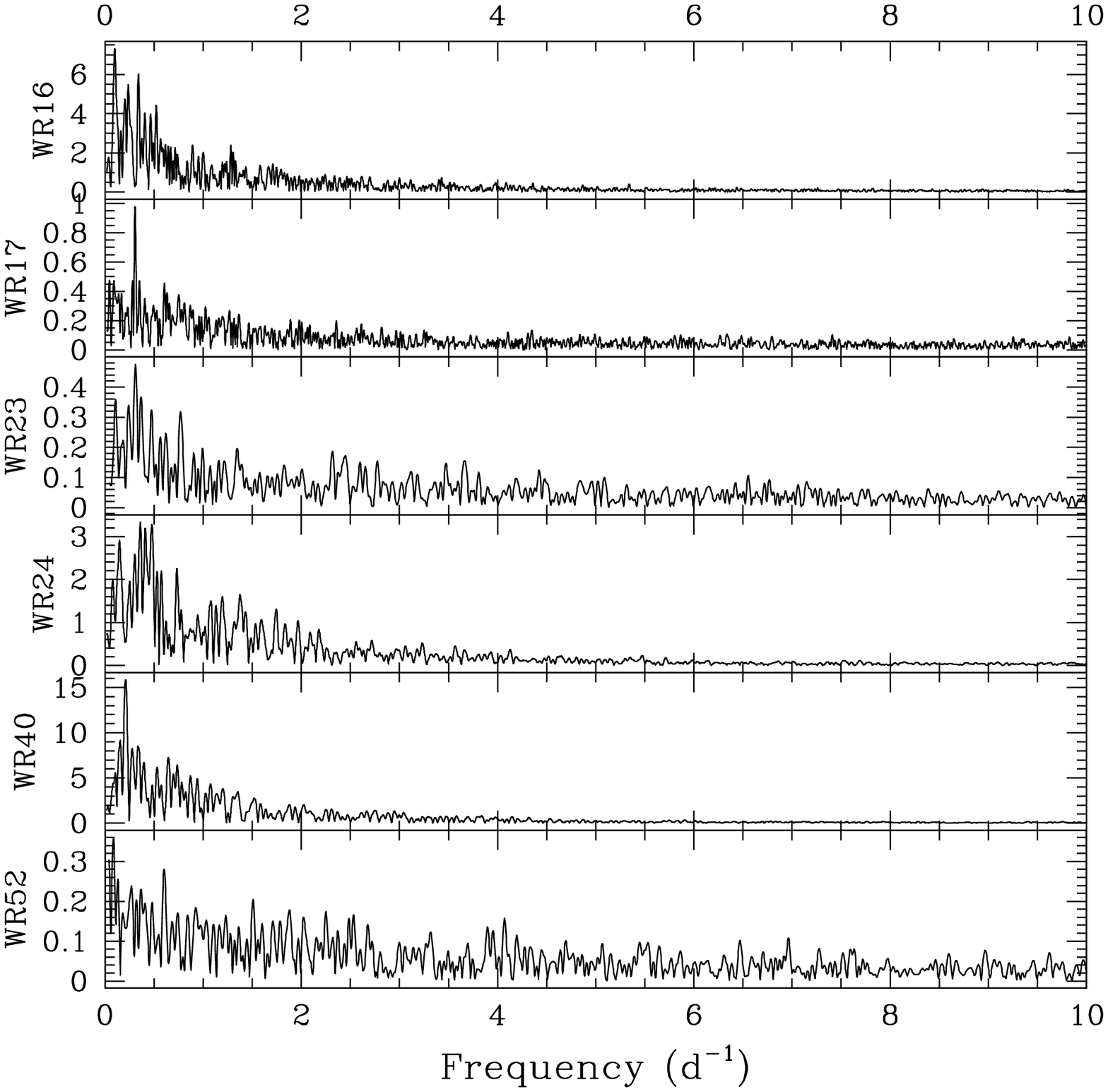}
\includegraphics[width=5.8cm]{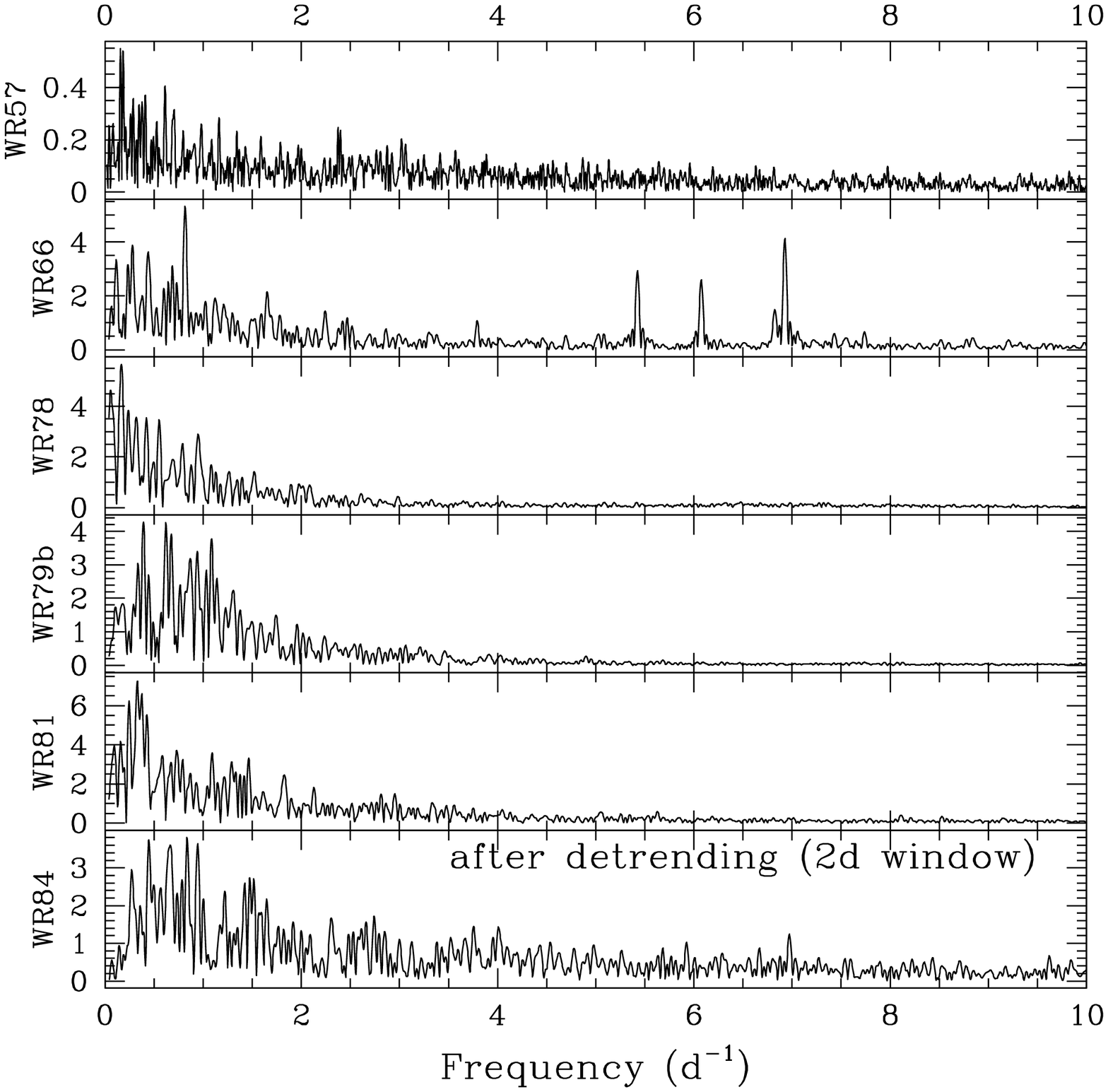}
\includegraphics[width=5.8cm]{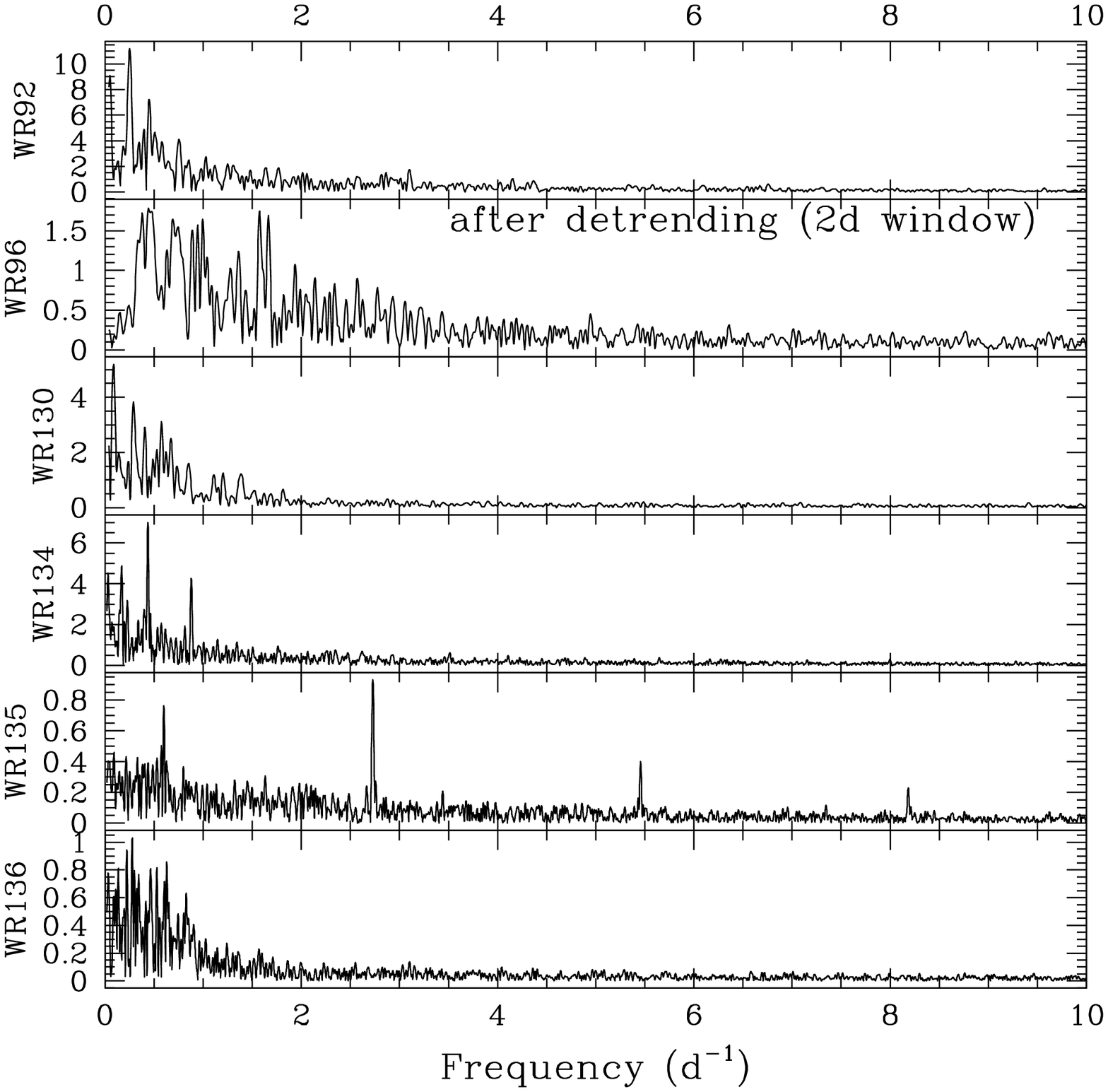}
\includegraphics[width=5.8cm]{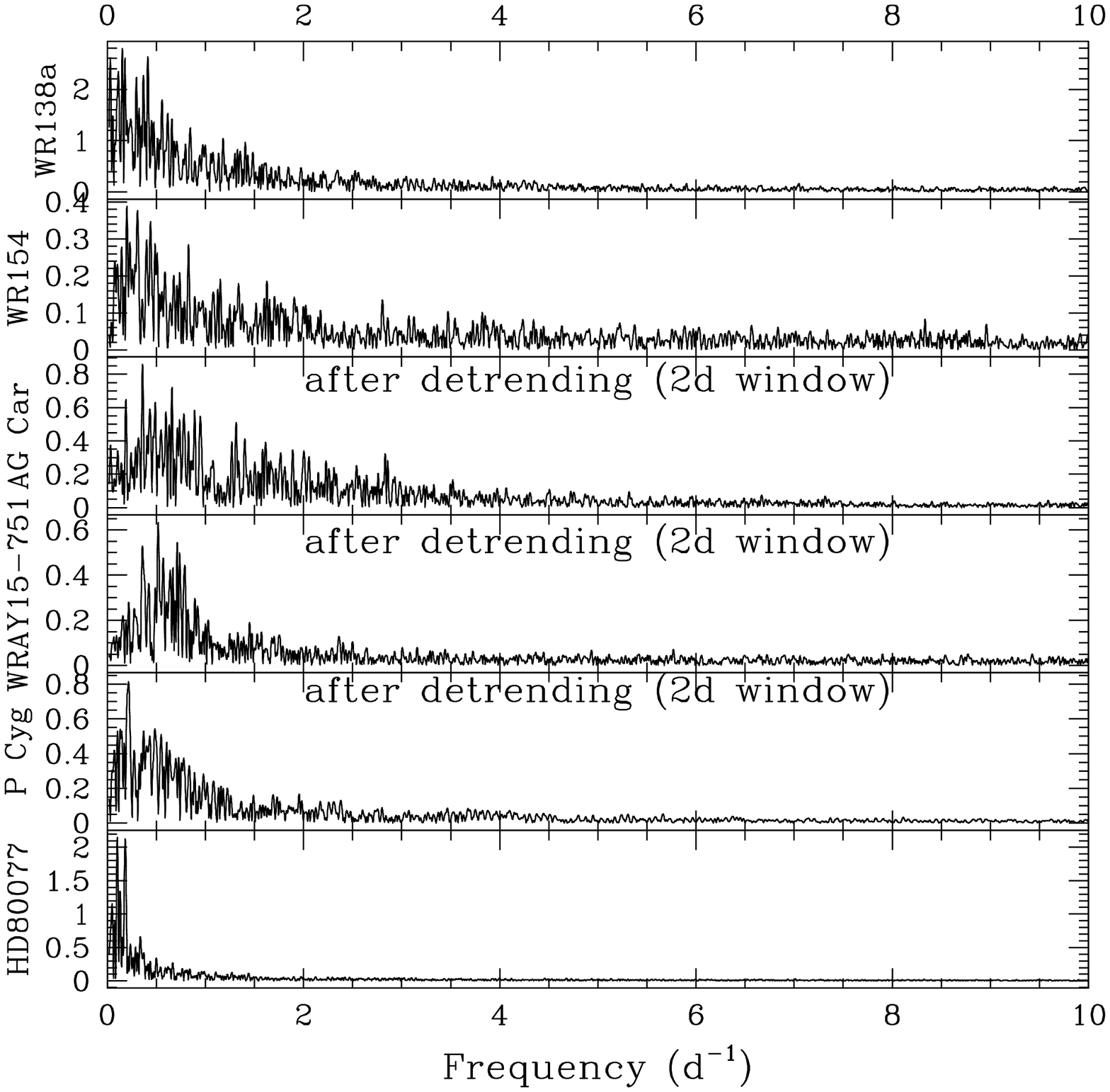}
\includegraphics[width=5.8cm]{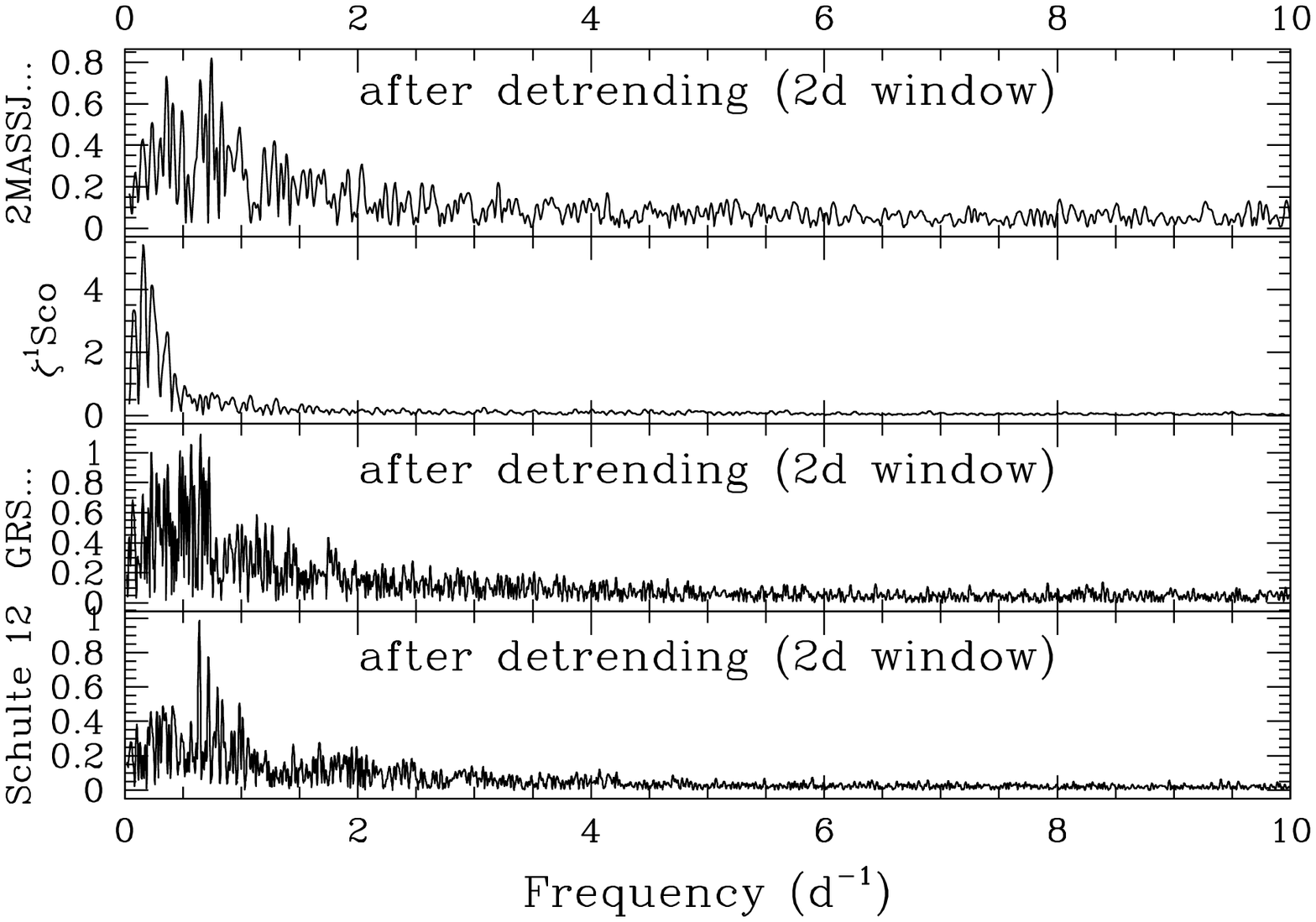}
  \end{center}
  \caption{Fourier periodograms associated to the lightcurves shown in Fig. \ref{lcwr}; the ordinate provides sinusoid amplitudes in mmag (i.e. $A$ if the sinusoid has the form $A \sin (2\pi\nu t+\phi)$). }
\label{fouwr}
\end{figure*}

Whatever the cadence, the raw fluxes were converted into magnitudes using $mag=-2.5\times \log(flux)$ and their mean was then subtracted. In several cases, the targets were observed over several sectors and the lightcurves were then combined. For eight stars (WR\,84 and 96, AG\,Car, WRAY\,15-751, P\,Cyg, 2MASS\,J16493770-4535592, GRS\,G079.29+00.46, and Schulte\,12), a slow, long-term trend is present in the photometry (Fig. \ref{detr}). As it could impair seeing short-term signals (the goals of this paper), it was determined using a 2\,d sliding window and then subtracted. In some cases, the beginning or end of observing windows show a steep upward/downward trend, due to an imperfect detrending, hence these parts were also discarded. The final lightcurves are shown in Fig. \ref{lcwr}.

\section{Results}
The \te\ lightcurves of our sample form a varied landscape. The scatter, for example, may take very different values. The lightcurves of LBVs and LBV candidates display scatter of 2\,mmag on average, but usually around long-term trends of much larger amplitudes. There is also a clear dichotomy between the WR subtypes. Nine out of the 12 WC-type stars have a scatter less than 2\,mmag, and the three remaining stars (WR\,81, 92, and 96) are all of the latest, WC9, type. In contrast, all but one (WR\,40) WN-type stars have scatter in the 2--14\,mmag range, without a clear distinction between early and late types. 

All lightcurves were analyzed using a modified Fourier algorithm adapted to uneven temporal samplings \citep{hmm,gos01,zec09}, as there is a small gap in the middle of each sector lightcurve. The periodograms are shown in Fig. \ref{fouwr}. In addition, to assess the evolution of the variability pattern, we derived the periodograms in sliding windows of 5\,d duration shifted by steps of 0.5\,d. These periodograms and the associated time-frequency diagrams clearly reveal the general presence of red noise, as well as of isolated peaks in a few cases.

\subsection{Red noise}
The gradual increase in power towards low frequencies in the periodograms, a so-called ``red noise'', now appears to be ubiquitous in massive O- and B-type stars \citep{blo11,rau19,bow19}. Such low-frequency variability could be produced by a combination of internal gravity waves excited at the interface between the convective core and the radiative envelope \citep{Rog13} or in a subsurface convection zone \citep{blo11}. Its presence has also been reported for a few WR stars (\citealt{gos90} and more recently \citealt{che11,dav12,ram19}) but its characteristics were not measured.

Since the pioneering work of \citet{har85} linked to ``solar noise'', several authors have used formulae of the type $a+ \frac{b}{1 + (c\,\nu)^d}$ to fit a mixed contribution of white and red noise components. There were however two independent approaches, both of which we consider below. In all cases, the fitting was performed using a Levenberg-Marquardt algorithm. As a few stars present isolated peaks (see next subsection), these peaks were excised from the periodograms before fitting. The fitting was performed up to 25\,d$^{-1}$ for stars with 30\,min cadence lightcurves and up to 360\,d$^{-1}$ for stars with 2\,min cadence lightcurves. The fitting began at 0.25\,d$^{-1}$ for stars where the lightcurves had been detrended. For WR\,24 and 40, lightcurves with 2\,min cadence are only available for one sector but we note that the fitted parameters are similar if considering only the 2\,min data or the combination of data from both sectors - only the former are presented in Table \ref{redn}. 

Whatever the formalism, estimating errors on the derived parameters is not an easy task. The diagonal elements of the variance-covariance matrix at best fit are well known to correspond to squared errors in the linear approximation: we will adopt that formalism, considering that it constitutes a first approximation in our non-linear case. However, it supposes that the fitting is done knowing the error on each input point while there is no formal error on the periodogram amplitudes. To overcome this problem, we may do the fitting without actual errors and consider, as is often done, that the best-fit reduced $\chi^2$ should amount to one. It should however be mentioned that the periodogram bins are correlated (a peak is spread over several bins and a peak at one place may depend on peaks at other places because of aliasing). The number of degrees of freedom in the frequency space thus does not correspond to the number of points in the fitted periodogram (which is arbitrarily chosen by the user) minus the number of fitted parameters (4). Instead, the actual number of independent frequencies in a periodogram, in the even sampling case, is half the number of points in the lightcurve. Therefore, parameter errors were assumed to be equal to the square root of the diagonal elements of the best-fit variance-covariance matrix, multiplied by the square root of $\chi^2({\rm best\,fit})/(0.5 \times N_{\rm data}-4)$ . The resulting errors are listed in Table \ref{redn}. Note that having performed the fitting in various conditions (different background region masks, fitting of one or two sectors' data,...) showed that these errors are probably slightly underestimated.

\subsubsection{Fitting amplitudes}

Brightness variations may be directly related to variations of physical parameters (e.g. $dL/L\propto dT/T$), and therefore constitute a prime interest of current asteroseismic studies of massive stars. Therefore, following several recent studies of massive stars \citep{blo11, rau19, bow19,bow20}, we may fit the periodogram amplitudes of our targets with:
\begin{equation}
  A(\nu) = C+ \frac{A_0}{1 + (2\,\pi\,\tau\,\nu)^{\gamma}}
  \label{eqrednoise}
\end{equation}
where $A_0$ is the red noise level at null frequency, $\tau$ the mean lifetime of the structures producing the red noise, $\gamma$ the slope of the linear decrease, and $C$ the white noise level.

The fit results are presented in the first part of Table \ref{redn}. For some stars (WR\,23, 52, 57, 78, 84, and 134 - for WR\,135 the situation is slightly better but $C$ remains slightly uncertain), the periodogram amplitudes continue to decrease at 25\,d$^{-1}$, i.e. the white noise plateau was not fully reached at that frequency hence we consider the fitting as incomplete and do not present its results in Table \ref{redn}. For a further star (HD\,80077), the best-fit appears somewhat inadequate by eye, with the drop to the white noise level occurring too abruptly; a smaller $\gamma$ coupled with a slightly larger $C$ provides a better-looking fit, although this ``eye-fitting'' remains qualitative. Therefore, we prefer not presenting the corresponding fit results.

\begin{table}
 \scriptsize 
  \caption{White noise and red noise parameters in \te\ data of WR stars and (c)LBVs, each group being ordered by R.A., for each fitting method; the error bars represent $\pm1\sigma$. The number of points in the \te\ lightcurves is provided in the last column of the bottom part of this Table. \label{redn}}
  \begin{tabular}{l c c c c c}
    \hline
\multicolumn{6}{l}{Amplitude fitting}\\    
    Star  & $C$  & $A_0$ & $\tau$ & $\gamma$ \\
          &(mmag)&(mmag) & (day)  &          \\
    \hline
WR1      &0.0151$\pm$0.0030 &2.970$\pm$0.091 &1.3359$\pm$0.0816 &1.30$\pm$0.04  \\
WR3$^*$	 &0.0189$\pm$0.0008 &0.572$\pm$0.007 &0.0140$\pm$0.0003 &1.84$\pm$0.04  \\
WR4$^*$	 &0.0147$\pm$0.0002 &0.212$\pm$0.005 &0.0763$\pm$0.0034 &1.29$\pm$0.03  \\
WR5      &0.0246$\pm$0.0049 &0.363$\pm$0.029 &0.1413$\pm$0.0197 &1.32$\pm$0.15  \\
WR7      &0.0323$\pm$0.0138 &7.170$\pm$1.250 &2.0701$\pm$0.6721 &1.02$\pm$0.08  \\
WR15     &0.0080$\pm$0.0018 &0.238$\pm$0.018 &0.3427$\pm$0.0543 &1.02$\pm$0.08  \\
WR16     &0.0382$\pm$0.0113 &3.225$\pm$0.114 &0.2313$\pm$0.0133 &1.76$\pm$0.10  \\
WR17     &0.0219$\pm$0.0017 &0.277$\pm$0.012 &0.1552$\pm$0.0114 &1.61$\pm$0.12  \\
WR24$^*$ &0.0091$\pm$0.0005 &1.921$\pm$0.020 &0.1454$\pm$0.0024 &1.87$\pm$0.03  \\
WR40$^*$ &0.0156$\pm$0.0017 &6.629$\pm$0.072 &0.1852$\pm$0.0032 &1.93$\pm$0.03  \\
WR66	 &0.0888$\pm$0.0130 &1.756$\pm$0.083 &0.1120$\pm$0.0071 &2.38$\pm$0.24  \\
WR79b$^*$&0.0122$\pm$0.0008 &1.782$\pm$0.018 &0.0919$\pm$0.0010 &3.85$\pm$0.12  \\
WR81     &0.0514$\pm$0.0201 &3.537$\pm$0.151 &0.1476$\pm$0.0097 &1.84$\pm$0.14  \\
WR92     &0.0659$\pm$0.0250 &4.587$\pm$0.252 &0.2163$\pm$0.0194 &1.69$\pm$0.14  \\
WR96$^\dagger$     &0.0647$\pm$0.0086 &1.145$\pm$0.063 &0.0976$\pm$0.0070 &2.08$\pm$0.19  \\
WR130    &0.0517$\pm$0.0089 &2.099$\pm$0.092 &0.2184$\pm$0.0135 &2.25$\pm$0.19  \\
WR135    &0.0036$\pm$0.0032:&0.299$\pm$0.015 &0.1199$\pm$0.0101 &1.20$\pm$0.09  \\
WR136    &0.0188$\pm$0.0014 &0.449$\pm$0.011 &0.1715$\pm$0.0054 &2.99$\pm$0.19  \\
WR138a   &0.0419$\pm$0.0048 &1.492$\pm$0.049 &0.2377$\pm$0.0128 &1.75$\pm$0.09  \\
WR154    &0.0154$\pm$0.0012 &0.168$\pm$0.008 &0.1497$\pm$0.0122 &1.58$\pm$0.12  \\
AG\,Car$^\dagger$                 &0.0091$\pm$0.0021 &0.378$\pm$0.019 &0.1188$\pm$0.0083 &1.80$\pm$0.12 \\
WRAY\,15-751$^\dagger$            &0.0218$\pm$0.0009 &0.251$\pm$0.009 &0.1564$\pm$0.0049 &4.17$\pm$0.37 \\
P\,Cyg$^\dagger$                  &0.0106$\pm$0.0009 &0.373$\pm$0.014 &0.1783$\pm$0.0078 &2.33$\pm$0.12 \\
{\tiny 2MASS\,J16493770-4535592$^\dagger$}&0.0478$\pm$0.0027 &0.353$\pm$0.024 &0.1164$\pm$0.0097 &2.36$\pm$0.27 \\
$\zeta^1$\,Sco$^*$               &0.0246$\pm$0.0007 &3.990$\pm$0.097 &0.7199$\pm$0.0317 &1.32$\pm$0.02 \\
GRS\,G079.29+00.46$^\dagger$      &0.0422$\pm$0.0023 &0.573$\pm$0.032 &0.1594$\pm$0.0115 &1.99$\pm$0.14 \\
Schulte\,12$^\dagger$             &0.0135$\pm$0.0018 &0.332$\pm$0.020 &0.1353$\pm$0.0109 &1.91$\pm$0.15 \\
\hline
\multicolumn{6}{l}{Amplitude$^2$ fitting}\\    
    Star  & $c$  & $a_0$ & $\tau$ & $\gamma$ & $N$\\
          &(mmag)&(mmag) & (day)  &          &    \\
    \hline
WR1      & 1.77$\pm$2.12:& 1.060$\pm$0.052 & 1.879$\pm$0.099 &2.11$\pm$0.12 & 3341 \\
WR4$^*$	 & 1.07$\pm$0.06 & 0.370$\pm$0.002 & 0.116$\pm$0.004 &1.81$\pm$0.05 & 14873\\
WR5      & 0.76$\pm$0.19 & 0.419$\pm$0.014 & 0.210$\pm$0.021 &2.28$\pm$0.31 & 1009 \\
WR7      & 1.19$\pm$3.92 & 2.016$\pm$0.211 & 1.219$\pm$0.153 &1.98$\pm$0.20 & 1069 \\
WR15     & 0.39$\pm$0.14 & 0.190$\pm$0.009 & 0.532$\pm$0.059 &1.74$\pm$0.16 & 1949 \\
WR16     & 3.58$\pm$4.08:& 2.717$\pm$0.084 & 0.346$\pm$0.019 &3.12$\pm$0.36 & 2293 \\
WR17     & 0.77$\pm$0.26 & 0.317$\pm$0.011 & 0.225$\pm$0.020 &2.52$\pm$0.35 & 2286 \\
WR24$^*$ & 2.78$\pm$1.28:& 1.871$\pm$0.013 & 0.215$\pm$0.003 &4.32$\pm$0.18 & 17575\\
WR40$^*$ & 3.40$\pm$15.4:& 6.383$\pm$0.083 & 0.321$\pm$0.010 &2.41$\pm$0.10 & 17592\\					     
WR57     & 0.51$\pm$0.26 & 0.284$\pm$0.009 & 0.224$\pm$0.027 &1.66$\pm$0.18 & 2483 \\
WR79b$^*$& 1.51$\pm$3.24 & 2.445$\pm$0.004 & 0.117$\pm$0.002 &5.94$\pm$0.41 & 12523\\				             
WR81     & 4.84$\pm$2.84:& 3.509$\pm$0.111 & 0.237$\pm$0.016 &3.25$\pm$0.47 & 1247 \\
WR92     & 5.38$\pm$5.01:& 3.939$\pm$0.175 & 0.301$\pm$0.021 &4.21$\pm$0.89 & 1257 \\
WR96$^\dagger$    & 1.24$\pm$1.26 & 1.569$\pm$0.042 & 0.131$\pm$0.010 &2.72$\pm$0.33 & 1167 \\
WR130    & 2.49$\pm$2.22:& 1.381$\pm$0.152 & 0.582$\pm$0.090 &2.83$\pm$0.74 & 1201 \\            
WR135    & 0.51$\pm$0.32 & 0.356$\pm$0.008 & 0.173$\pm$0.013 &2.07$\pm$0.20 & 2370 \\
WR136    & 0.65$\pm$0.51 & 0.457$\pm$0.009 & 0.208$\pm$0.007 &5.07$\pm$0.64 & 2364 \\
WR138a   & 1.28$\pm$1.89 & 1.265$\pm$0.035 & 0.341$\pm$0.020 &2.54$\pm$0.23 & 2076 \\
WR154    & 0.58$\pm$0.09 & 0.191$\pm$0.005 & 0.221$\pm$0.014 &2.73$\pm$0.29 & 2161 \\
AG\,Car$^\dagger$           &0.49$\pm$0.49 &0.452$\pm$0.013 &0.165$\pm$0.011 & 2.74$\pm$0.29  & 2400 \\
WRAY\,15-751$^\dagger$      &0.80$\pm$0.15 &0.300$\pm$0.010 &0.185$\pm$0.011 & 3.65$\pm$0.48  & 2072 \\
{\tiny 2MASS\,J16493770-4535592$^\dagger$}&1.04$\pm$0.16 &0.463$\pm$0.013 &0.140$\pm$0.008 & 4.38$\pm$0.74 & 1104 \\
$\zeta^1$\,Sco$^*$         &1.85$\pm$2.30 &1.892$\pm$0.025 &0.502$\pm$0.008 & 4.02$\pm$0.18  &12434\\
GRS\,G079.29+00.46$^\dagger$&1.61$\pm$0.27 &0.584$\pm$0.005 &0.183$\pm$0.007 & 4.63$\pm$0.54  & 2343\\
Schulte\,12$^\dagger$       &0.71$\pm$0.30 &0.367$\pm$0.003 &0.160$\pm$0.009 & 4.48$\pm$0.79  & 2027\\
    \hline
  \end{tabular}

{\scriptsize $^*$ indicates a fitting performed up to 360\,d$^{-1}$, $^\dagger$ a fitting beginning at 0.25\,d$^{-1}$, : slight overestimates. } 
\end{table}

\begin{figure}
  \begin{center}
\includegraphics[width=8cm]{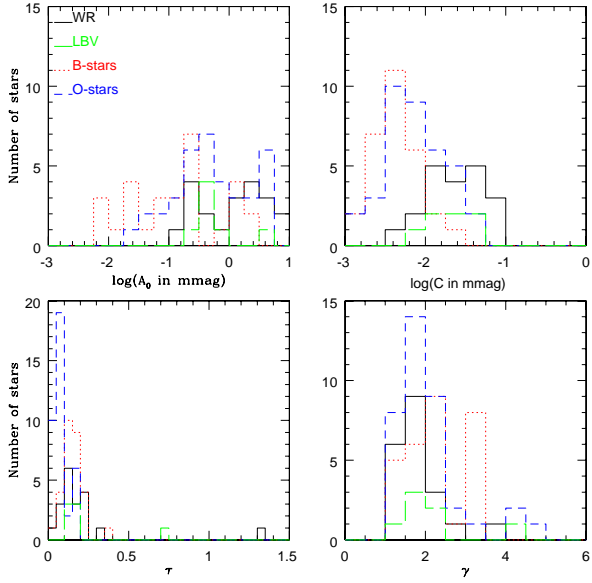}
  \end{center}
  \caption{Histograms of the white+red noise parameters fitted to periodograms of lightcurves from O-stars (dashed blue line, \citealt{bow20}), B-stars (red dotted line, \citealt{bow20}), WR stars (black solid line, this work), and (c)LBVs (green long dashed line, Sect. 3.1.1 of this work).}
\label{comparn}
\end{figure}

\begin{figure}
  \begin{center}
\includegraphics[width=8cm]{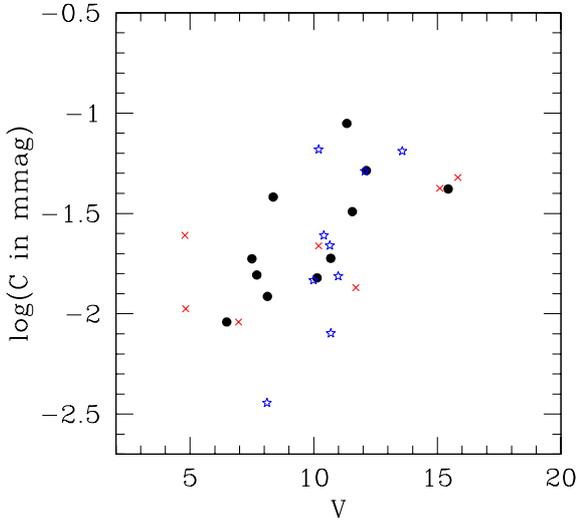}
  \end{center}
  \caption{Comparison between the magnitude of the targets and the fitted white noise value (Sect. 3.1.1, top of Table \ref{redn}).}
\label{magnoise}
\end{figure}

\begin{figure*}
  \begin{center}
\includegraphics[width=8cm]{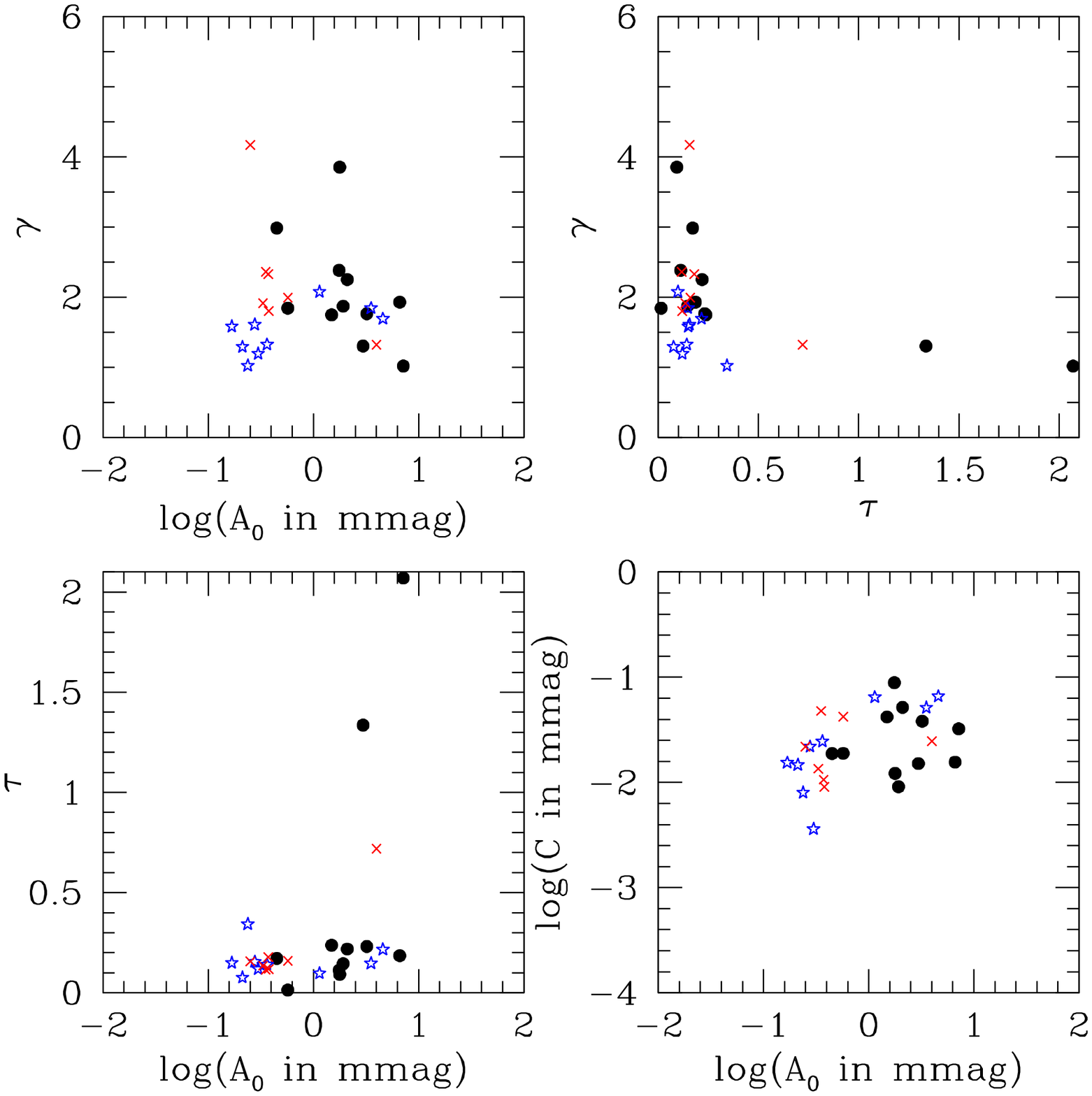}
\includegraphics[width=8cm]{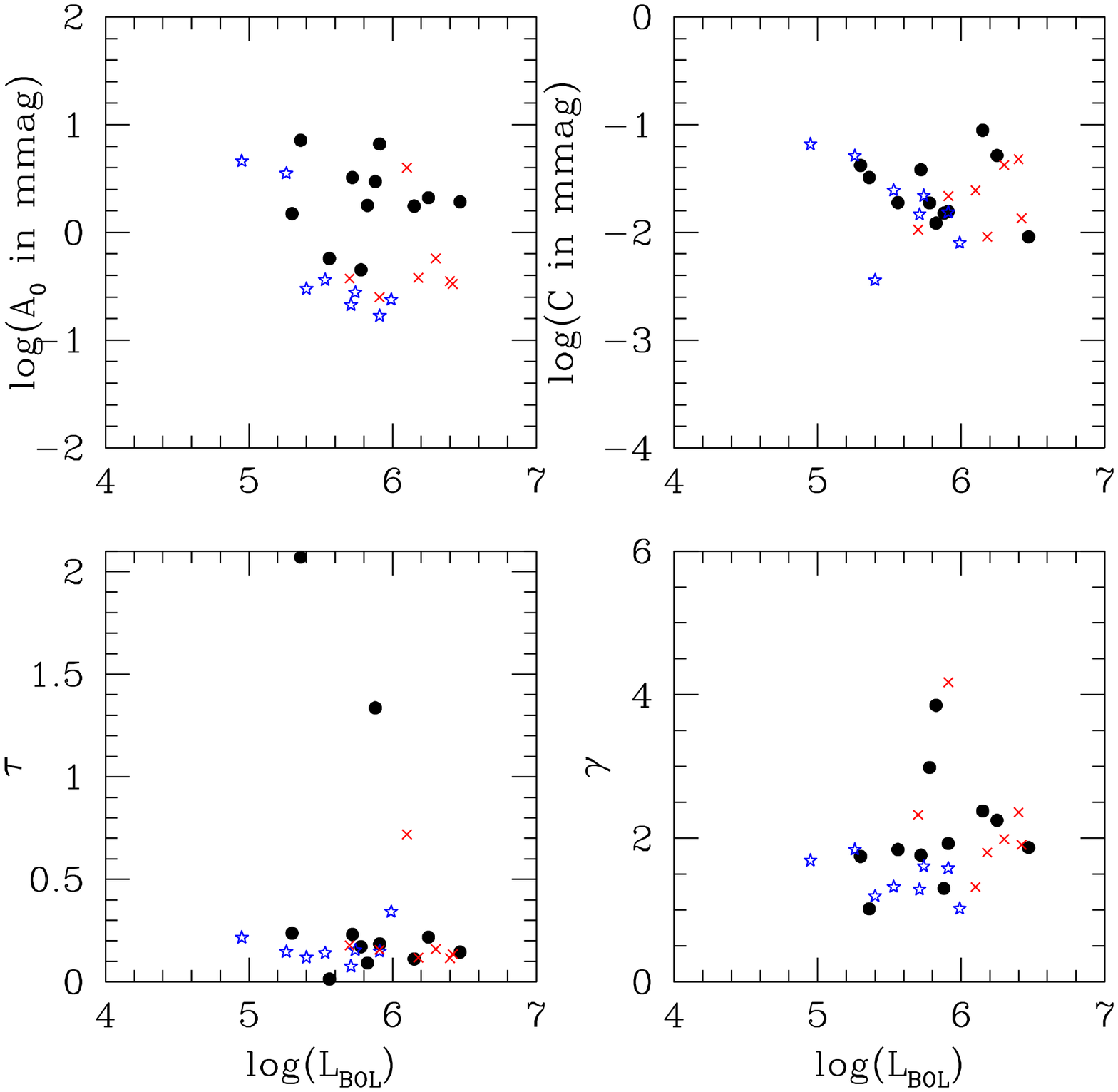}
  \end{center}
  \caption{Relationships between red+white noise parameters (Sect. 3.1.1, top of Table \ref{redn}) and bolometric luminosities. Black dots indicate WN stars, blue stars WC stars, and red crosses (c)LBVs.}
\label{paramrn}
\end{figure*}

Figure \ref{comparn} compares our red noise parameters for 20 single WR stars and 7 LBVs and candidates to those found for 37 O-stars and 29 early B-stars by \citet{bow20}. As can be seen, the $\gamma$ values appear similar for all stars. $A_0$ and $\tau$ may be slightly larger for evolved stars but the difference is small. The derived similarity in red noise parameters implies that the physical phenomenon responsible for this variability displays similar features in all massive stars. However, in OB-stars, one can directly see the photosphere whereas in WR stars, the hydrostatic surfaces of the stars remain hidden, with the unity optical depth residing inside the wind. On the one hand, the red noise in OB-stars is often considered as the result of internal gravity waves excited by turbulence (either in the convective core or in the near-surface convective layers, see e.g. \citealt{bow20}). On the other hand, \citet{ram19} demonstrated that a variability similar to that of WR\,40 could be produced by a stochastically clumped wind. However, the wind mass loss is directly proportional to the surface radiation flux \citep{luc93,gra17} and if it changes due e.g. to pulsations, then changes in the overlying wind are expected. In particular, a link between wind clumping and perturbations at the level of the hydrostatic radius would not be surprising and our result may possibly be an indirect evidence for it. 

In parallel, $C$ clearly varies, with B-stars presenting the lowest white noise levels and WRs and (c)LBVs the largest ones - on average, the difference between them amounts to one dex. However, the WR stars of our sample are intrinsically much brighter and/or hotter (and the (c)LBVs brighter) than the OB-stars of \citet{bow20}. Comparing the bright O-type giants and supergiants with the faintest and coolest WR-stars (WR\,16, 40, 79b, 81, 92, and 138a), the difference appears less extreme ($C$ of 4--38\,$\mu$mag vs 12--66\,$\mu$mag). We thus probably observe a continuous trend in white noise levels, driven by temperature and/or luminosity and/or mass-loss effects. A last effect should also be taken into account: the influence of photon noise. Indeed, the intrinsic white noise of the stellar emission is mixed with the photon noise that depends on the apparent magnitude and instrumental sensitivity. In this context, it is interesting to note that lower $C$ values can be found for the visually brighter objects while the faintest targets do not display small $C$ values (Fig. \ref{magnoise}).

Finally, we checked for correlations between the fitted noise parameters and between them and the stellar properties (Fig. \ref{paramrn}). This was done for the whole WR sample, but also for WN or WC stars separately. Examining first the noise parameters themselves, the largest Pearson correlation coefficient is found between the noise levels $A_0$ and $C$, especially for WC stars (0.5 for the whole sample but 0.8 for WC). When examining the link with stellar properties, negative correlation coefficients are detected for WC stars between stellar luminosities and both noise levels (coefficients of $\sim$\,--0.9 for $A_0$ and --0.6 for $C$), i.e. less noise for intrinsically more luminous stars. There is also a negative correlation (coefficients of --0.6 for all WR groups) between $\gamma$ and stellar temperatures, i.e. slower transitions red$\rightarrow$white noise for hotter stars. None of the coefficients are extremely significant, though. In contrast, for OB-stars, \citet{bow20} found that the red noise level and lifetime $\tau$ ($=1/(2\pi\nu_{\rm char})$) decreased towards the ZAMS (i.e. higher temperatures and lower luminosities). No obvious correlation appears in the plots for (c)LBVs but given the small number of these targets, it is more difficult to draw general conclusions for them.

In summary, while the red noise parameters of evolved massive stars appear overall similar to those of OB-stars, their relation to stellar properties seems different, possibly more complex. 

\begin{figure*}
  \begin{center}
\includegraphics[width=6.7cm]{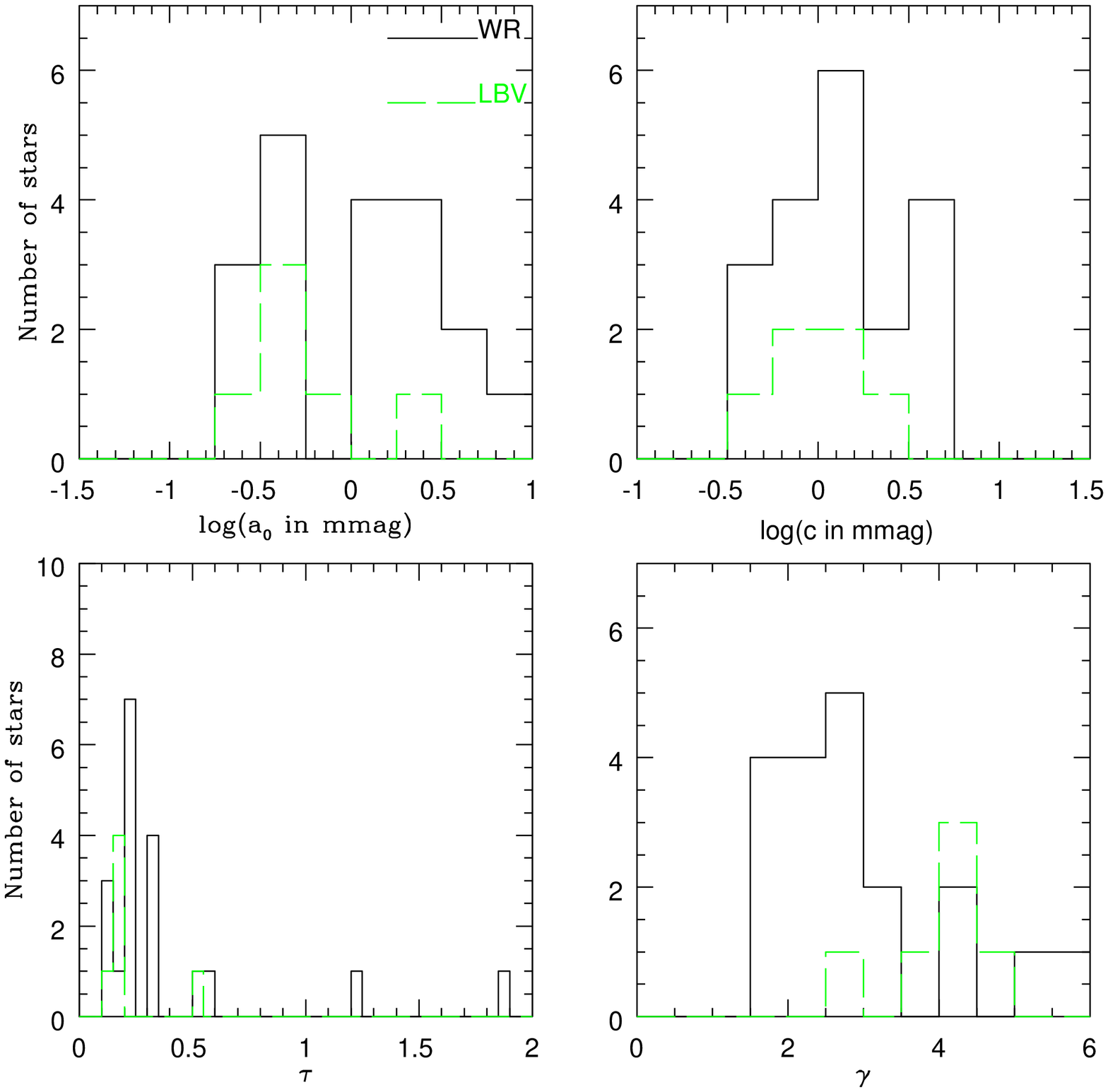}
\includegraphics[width=6.7cm]{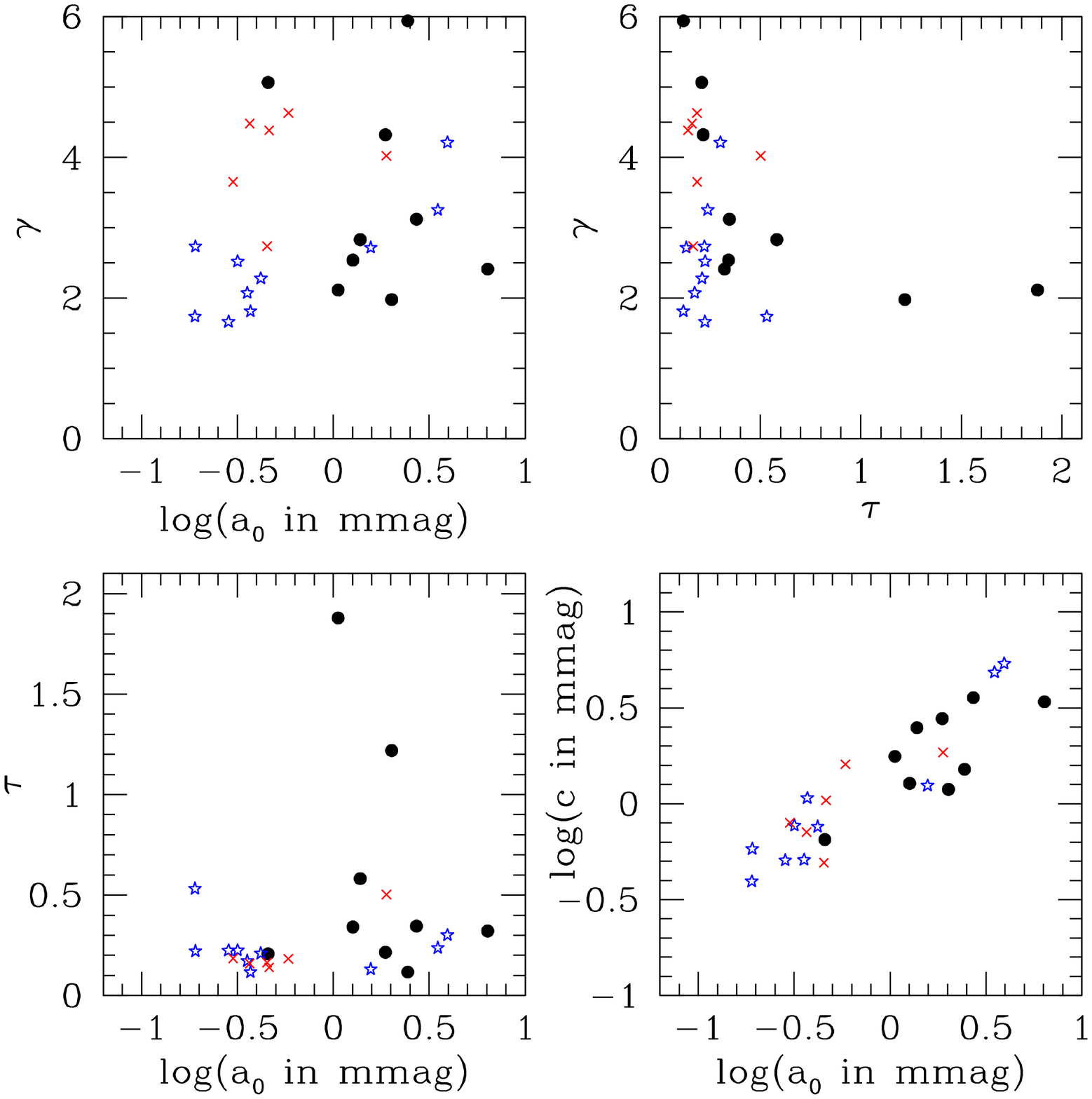}
\includegraphics[width=3.65cm,bb=280 145 590 715,clip]{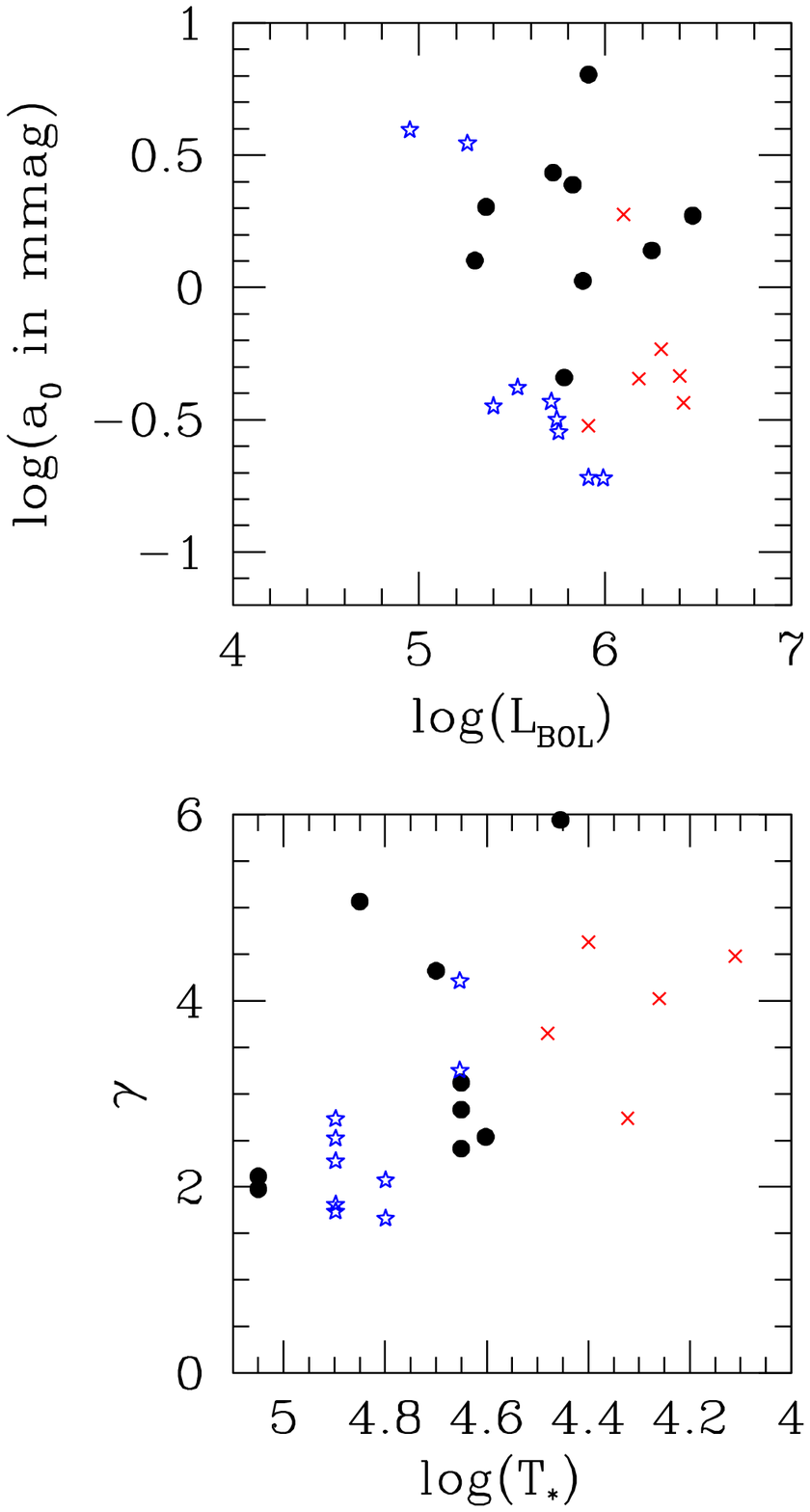}
  \end{center}
  \caption{Same as Figs. \ref{comparn} and \ref{paramrn}, but for the squared amplitude fitting.}
\label{psd}
\end{figure*}

\subsubsection{Fitting squared amplitudes}
In the case of data generated by stochastic processes, observed time-series are stochastic entities and so are the associated Fourier and Power Spectra. Following the prescriptions of \citet{dee75}, the combination of stochastic signals (white noise WN, red noise RN) must be performed through the addition of their Power Spectra, i.e. $P(\nu) \, = \, P_{\mathrm{WN}} \, + \, P_{\mathrm{RN}}(\nu)$. In other words, stochastic processes sum up quadratically, not linearly, hence a more correct fitting of stochastic processes should actually consider Power Spectra instead of Amplitudes. 

A linear process, such as a damped oscillator, excited by a white noise (a general form for a stochastic process) generates time-series whose Power Spectra are proportional to Lorentzian functions, leading to the general expression for the red noise process already mentioned above. This was the method adopted for granulation studies \citep{har85,kal14}, but also in a cataclysmic variable analysis \citep{Stanishev} and in one recent massive star analysis \citep{bowcorot}.

Following \citet{kal14}, we therefore consider the sum of the powers:
\begin{equation}
  P(\nu) = c^2+ \frac{2\,\pi\,\tau\,\xi\,P_0}{1 + (2\,\pi\,\tau\,\nu)^{\gamma}}
\end{equation}
where $\tau$, and $\gamma$ have similar meanings as before, $c^2$ and $P_0$ are the white and red noise strengths, and $\xi$ is a normalization factor \citep{kal14}. Indeed, $\int_0^{\infty} \frac{dx}{1 + x^{\gamma}}=\frac{\pi}{\gamma \sin(\pi/\gamma)}$ for $\gamma>1$: to normalize the integral, a factor $\xi=\frac{\gamma}{\pi}\sin(\frac{\pi}{\gamma})$ must then be used. It ensures that $P_0$ in equation (2) above truly represents all power due to red noise.

As defined by \citet{dee75}, \citet{sca82}, and \citet{hmm}, power and amplitude are related by $P(\nu) \, = \, \frac{N}{4} A^2(\nu)$ (see e.g.\ Sect.\ II.a of \citealt{sca82}) where $N$ is the number of data points in the lightcurve. The above equation may then be rewritten as:
\begin{equation}
  A^2(\nu) = \frac{4}{N} c^2+ \frac{2\,\pi\,\tau\,\xi\,a_0^2}{1 + (2\,\pi\,\tau\,\nu)^{\gamma}}
  \label{eqrednoise2}
\end{equation}
where $a_0$ is the red noise amplitude.

We thus performed once again the fits, now using the formalism of Eq. (3). The achieved fittings were however of lower quality than in the amplitude-fitting case. In particular, we note that the white noise level is often overestimated: its worse fitting is probably linked to the amplitude squaring which makes its contribution smaller, the fitting procedure then favoring the achievement of a good fit at low frequencies. When the white noise plateau is not reached, its correct determination becomes even more difficult, of course. The bottom part of Table \ref{redn} thus only lists the fitting results which appear reasonably secure.

As for the previous fitting, we investigated the correlations between parameters derived for WR stars (bottom part of Table \ref{redn} and right part of Fig. \ref{psd}). Immediately, a strong correlation (Pearson coefficient $>0.9$) is found between the two noise levels, confirming the doubts on the white noise level determination. Regarding other parameters, the $A^2$ fitting brings a confirmation to results found in previous subsection: there is an anticorrelation between bolometric luminosities of WC stars and the red noise levels (coefficient of $-0.9$) as well as an anticorrelation between the slopes and the stellar temperatures (coefficients of $-0.5$ to $-0.7$).

Comparing our results for evolved massive stars with typical parameters of OB-stars appears rather difficult as there is only one published study, that of \citet{bowcorot}. They achieved reasonable fittings for only 5 O-stars and 2 early B-stars, although with a different definition of power (that chosen by \citealt{deg10}). While this prohibits the comparison of noise levels, the $\tau$ and $\gamma$ values can still be directly compared (Table \ref{redn} and left part of Fig. \ref{psd}). In \citet{bowcorot}, the slopes $\gamma$ of the 7 OB-stars were found to be between 1.8 and 3.3 in all but one case, a range in line with those found here for WR stars. Furthermore, the lifetimes $\tau$ ranged between 0.06 and 0.14\,d$^{-1}$ for OB-stars, on average somewhat smaller values than for WR stars (where most values lie between 0.1 and 0.35\,d$^{-1}$). Both conclusions are fully in line with what was derived from the fitting of the amplitude spectra.

\subsection{Coherent frequencies}
 \begin{figure}
  \begin{center}
\includegraphics[width=8cm]{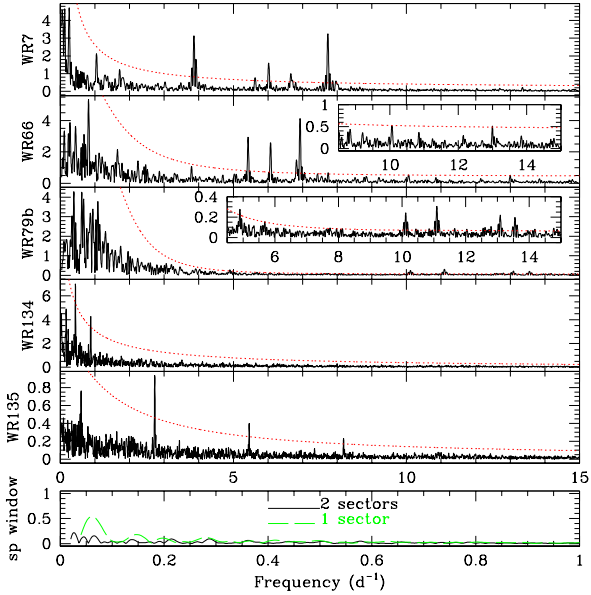}
  \end{center}
  \caption{Fourier periodograms showing isolated peaks, with 5 times the red+white noise level (Sect. 3.1.1) marked by the red dotted line. The y-axis provides amplitudes in mmag. The bottom panel provides the typical spectral window for observations in one sector (which is the case of the first three panels) and two sectors (which is the case of the next two panels).}
\label{freq}
\end{figure}

While no (c)LBV shows this feature, five WR stars (i.e. a fifth of our WR sample) display isolated peaks, reminiscent of coherent variability. These stars are WR\,7, 66, 79b, 134, and 135 (Fig. \ref{freq}). Assessing their significance requires to take the presence of red noise into account. Indeed, an overall significance level, derived e.g. from the data scatter \citep{mah11} or from the mean level outside strong peaks, is valid for all frequencies hence can only be used when the ``background'' periodogram level does not change with frequency. Therefore, we compared the peak amplitudes to five times the best-fit red+white noise model computed in the Sect. 3.1.1\footnote{The $A^2$ fitting was not used as its quality was lower, especially at high frequencies. However, we may note that, if adopted, we would reach the same list of significant signals.} (this corresponds to a signal-to-noise ratio of five at a predefined frequency, as often recommended - see e.g. \citealt{bar15}); no iterative cleaning was performed. The outstanding frequencies are listed in Table \ref{ftab}. 

\begin{figure*}
  \begin{center}
\includegraphics[width=5.8cm]{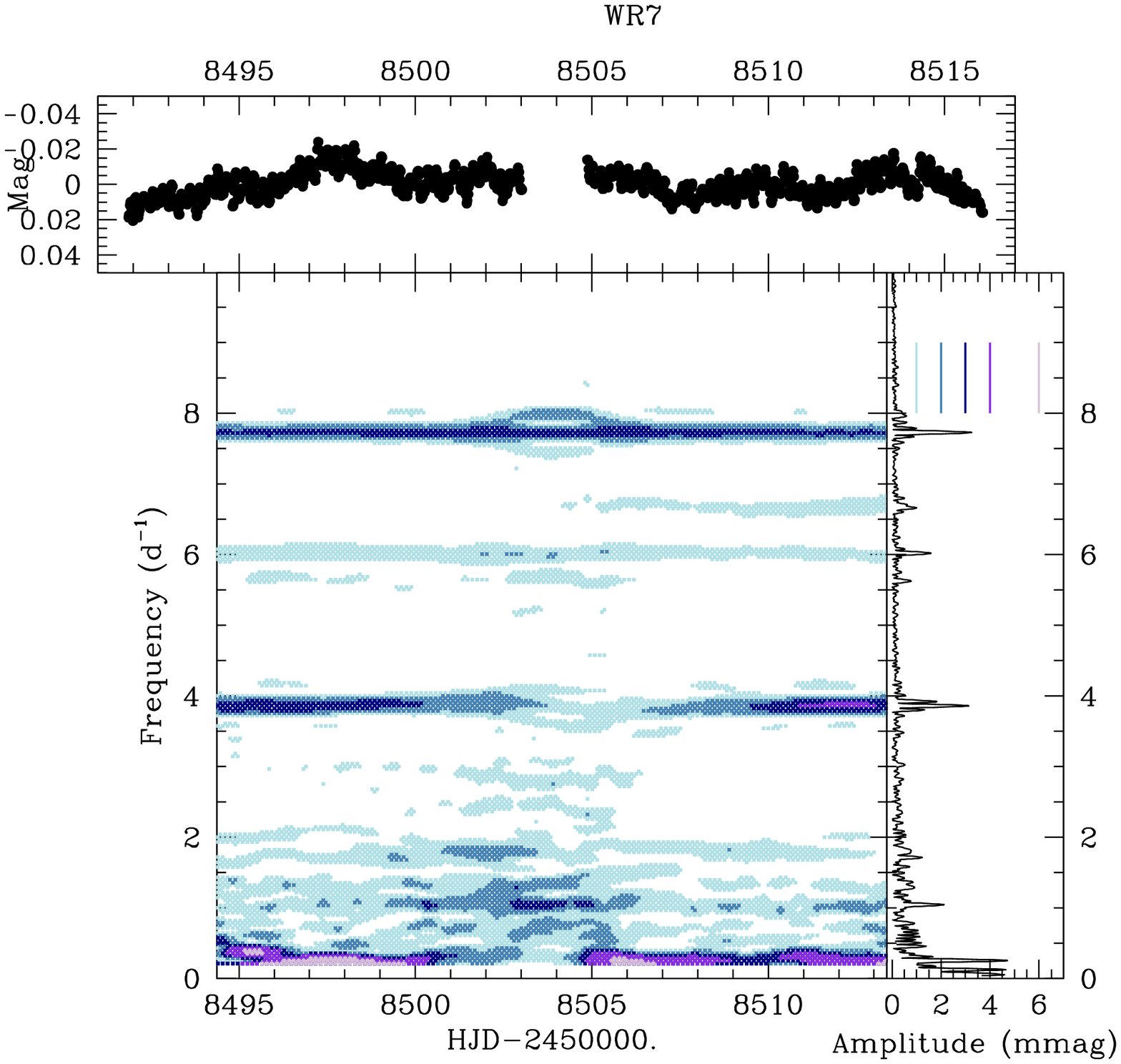}
\includegraphics[width=5.8cm]{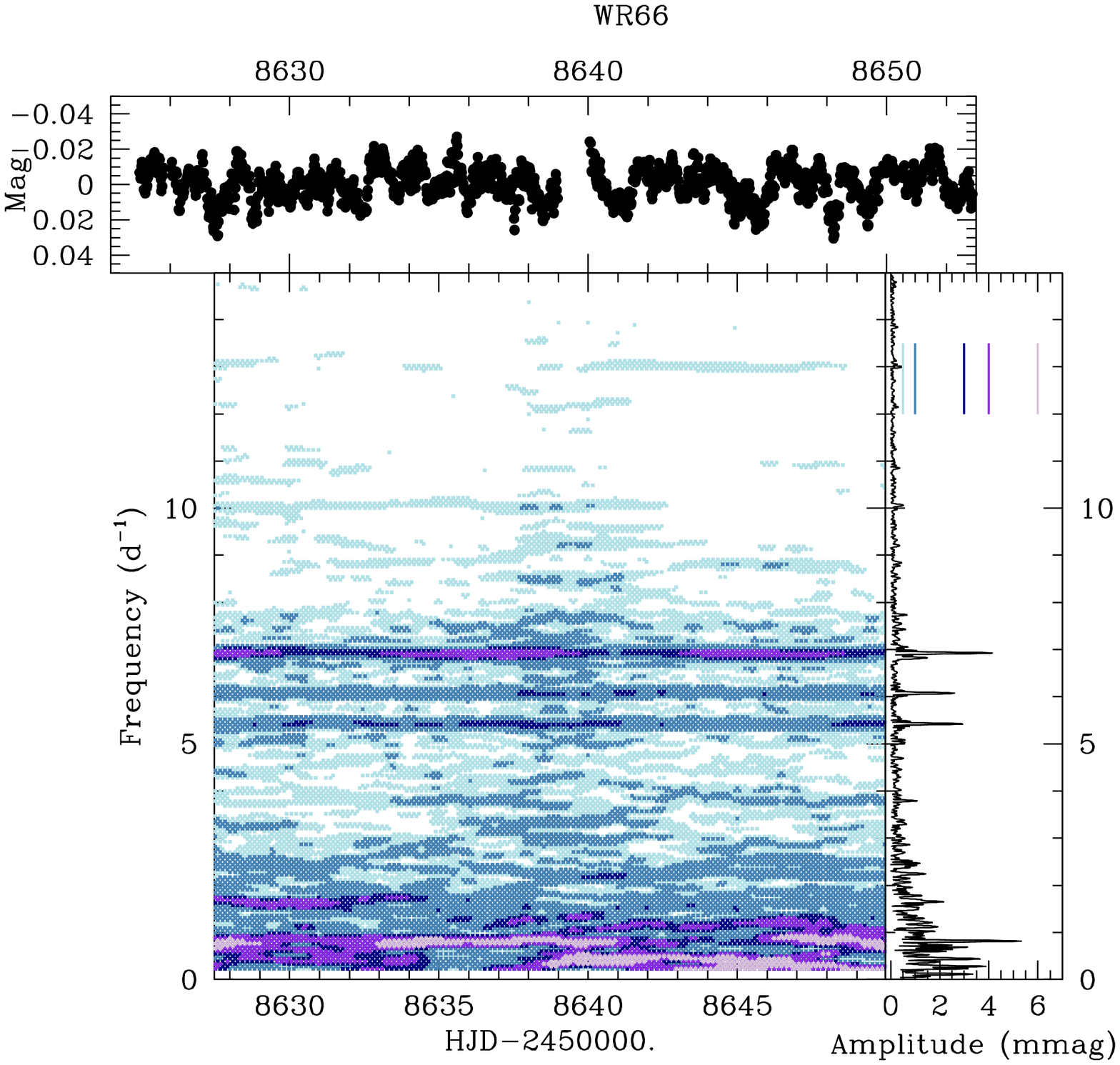}
\includegraphics[width=5.8cm]{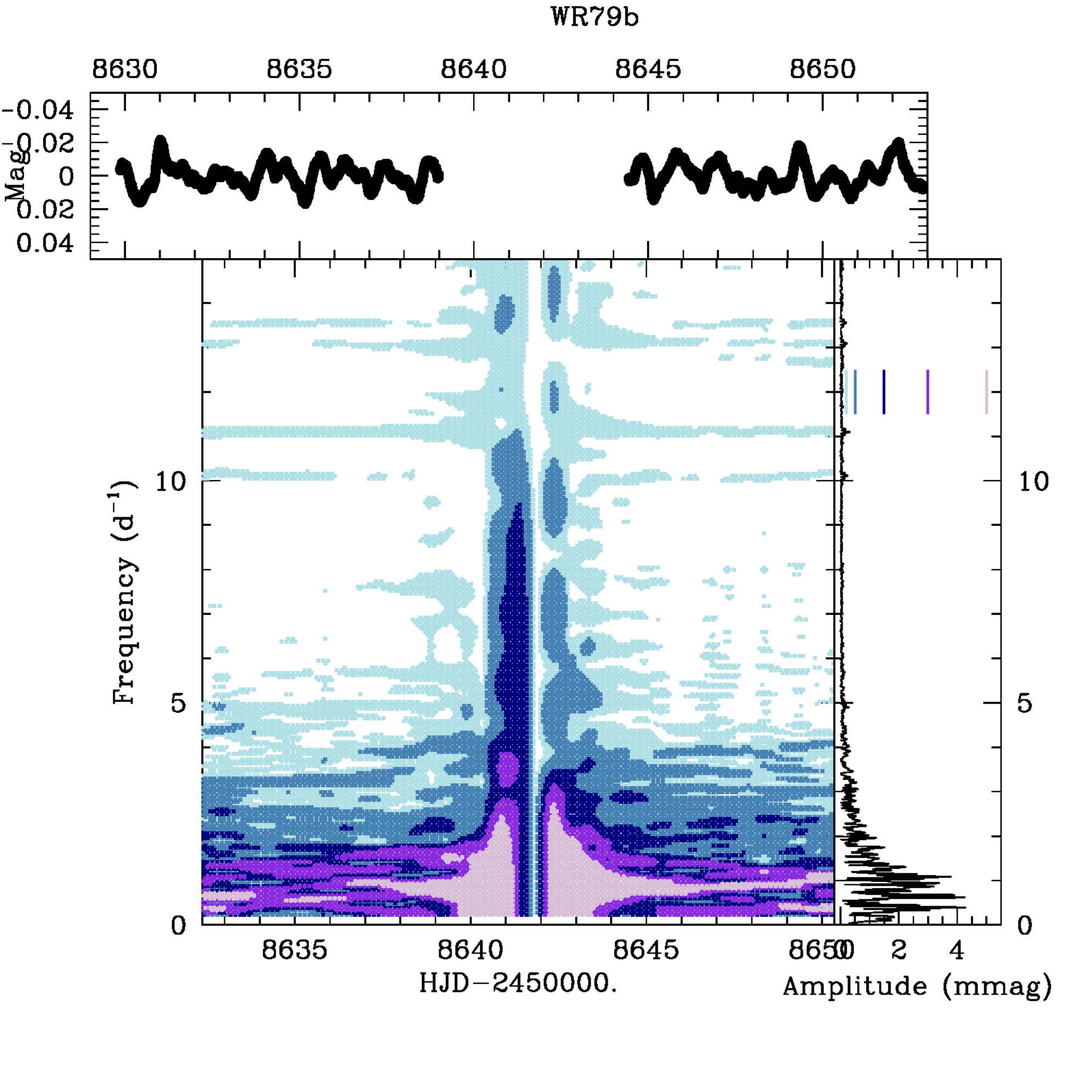}
\includegraphics[width=5.8cm]{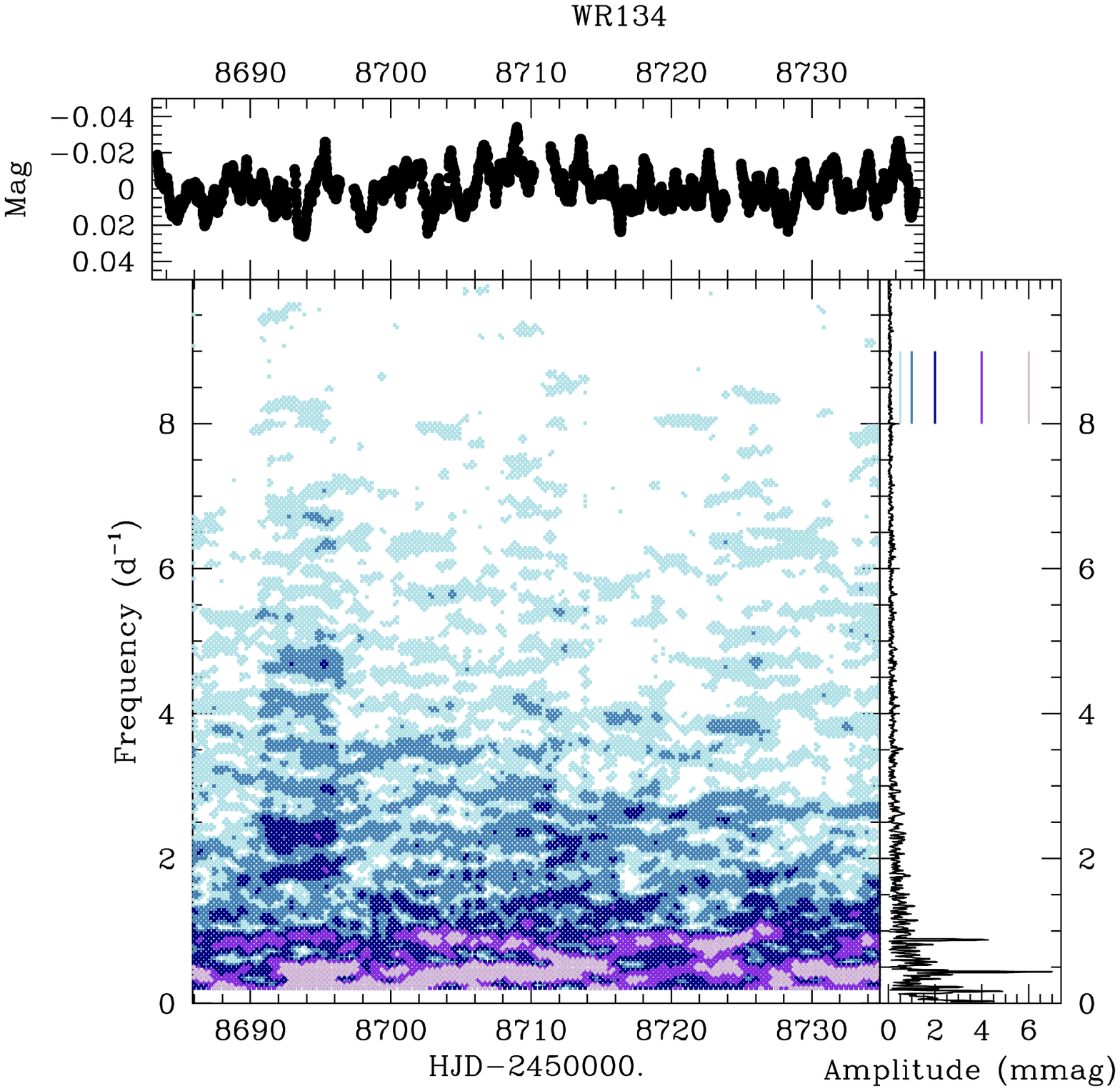}
\includegraphics[width=5.8cm]{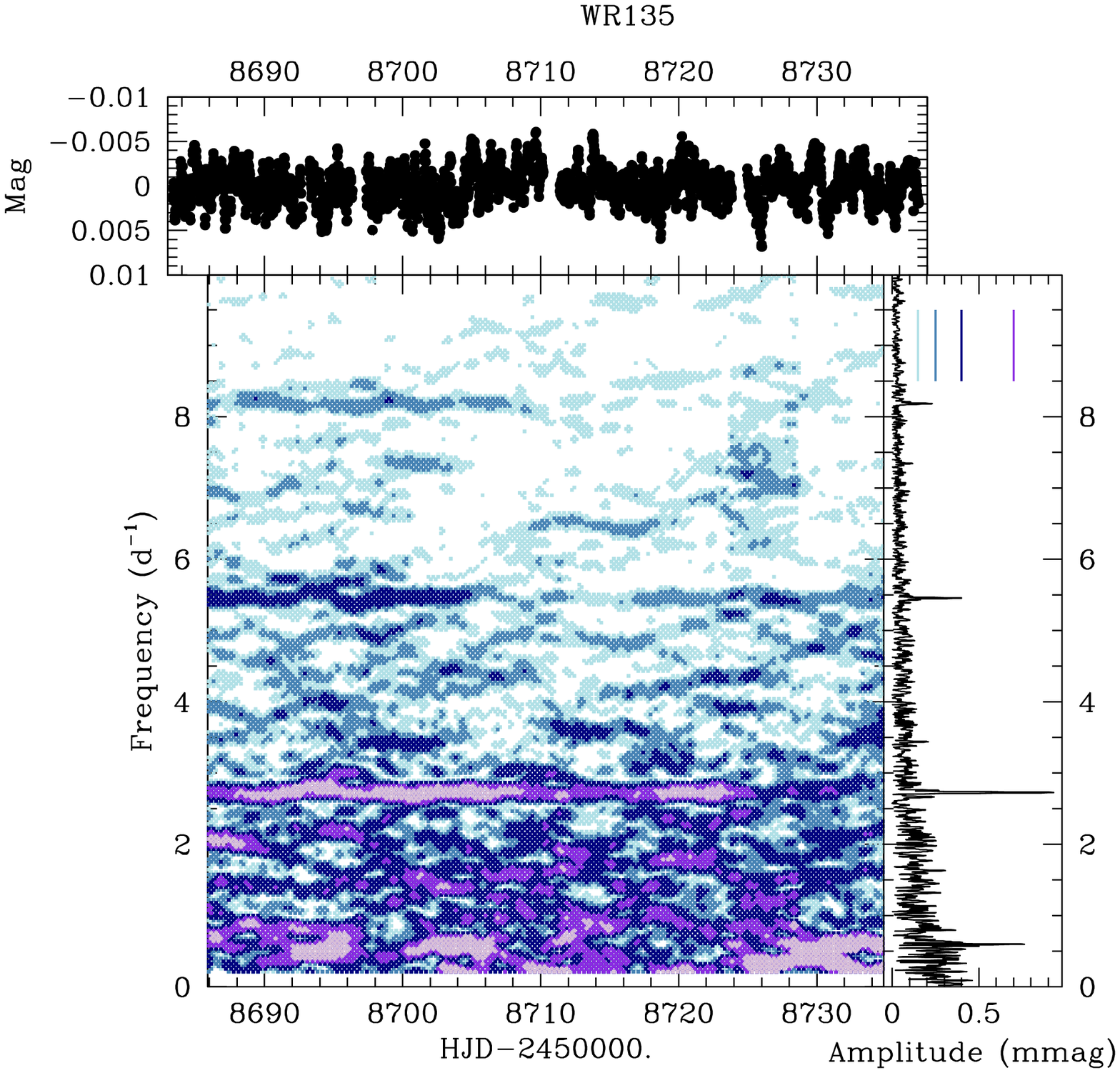}
\includegraphics[width=5.8cm]{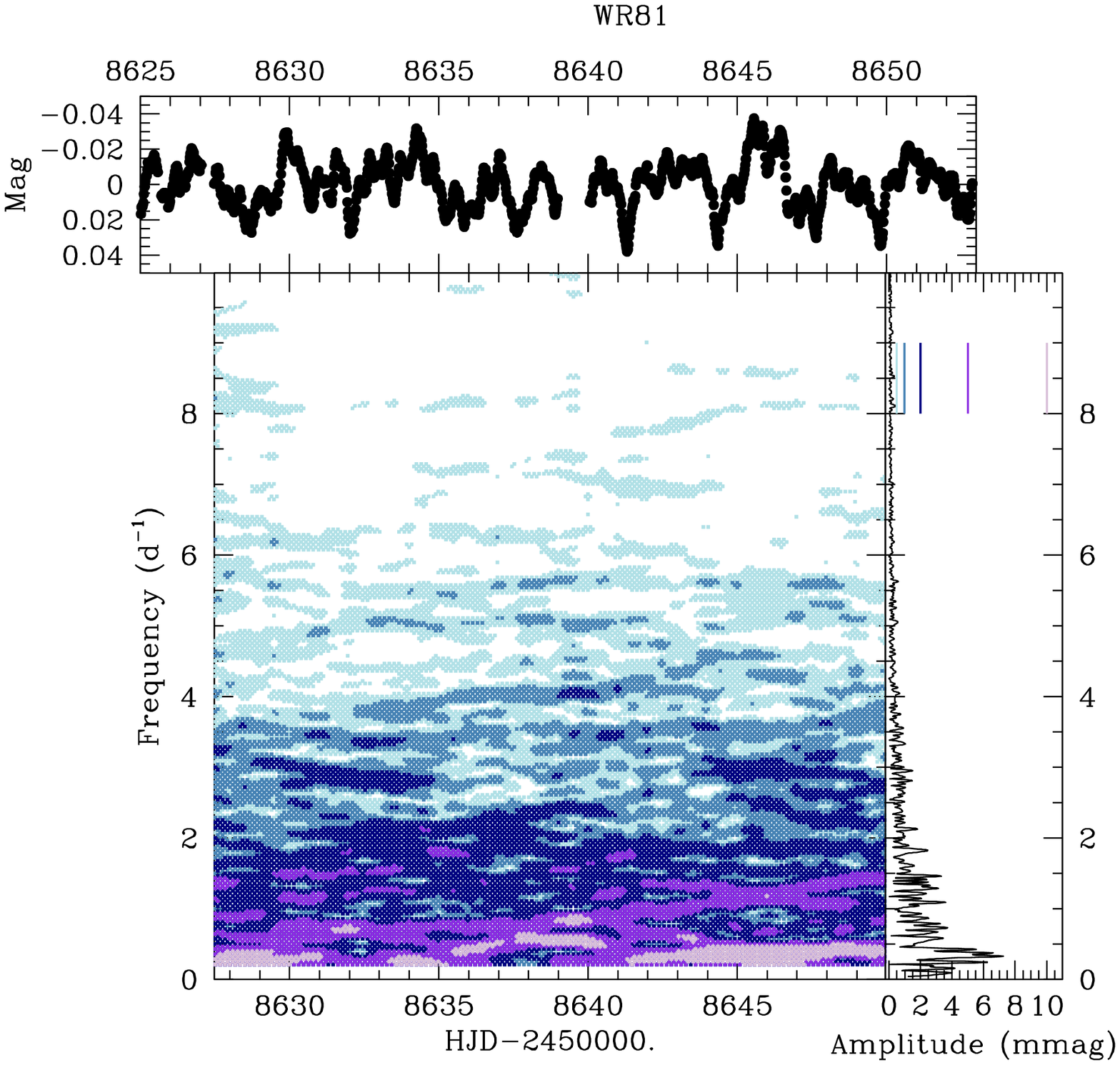}
  \end{center}
  \caption{Time-frequency diagrams for the five WRs presenting isolated peaks (WR\,7, 66, 79b, 134, and 135), compared to the case of WR\,81, whose periodogram only displays red+white noise. The dates in abscissa correspond to the mid-point of the temporal window used for calculating the periodogram. The lightcurve is displayed on top, whilst the periodogram for the full dataset is shown on the right. Short colored lines in the periodogram panels indicate the thresholds used for the colour scheme of time-frequency diagrams. }
\label{tfdiag2}
\end{figure*}

\begin{table*}
  \caption{Detected frequencies.  \label{ftab}}
  \begin{tabular}{lcc}
    \hline
    Star  & Sp.type & $\nu$ in d$^{-1}$ (ampl. in mmag) \\
    \hline
WR\,7   &WN4s	& 3.860 (3.14), 5.628 (0.77), 6.020 (1.60), 6.660 (1.00), 7.728$^*$(3.25) \\
WR\,66  &WN8(h)	& 5.424 (2.94), 6.076 (2.59), 6.928 (4.13), 10.062 (0.52), 12.996 (0.51) \\
WR\,79b &WN9ha	& 4.900 (0.28), 10.116 (0.24), 11.096 (0.31), 13.088 (0.22), 13.552 (0.20) \\
WR\,134 &WN6s	& 0.438 (7.01), 0.880$^*$(4.26) \\
WR\,135 &WC8	& 2.729 (0.93), 5.456$^*$ (0.40), 8.184$^*$(0.23) \\
    \hline
  \end{tabular}

{\scriptsize $^*$ indicates harmonics. Peak widths are 0.04\,d$^{-1}$ for the first three stars (one sector observations) and 0.02\,d$^{-1}$ for the last two (two sector observations); the errors on the peak frequencies are a fraction of that value (typically one tenth). Peak amplitudes from the periodograms are quoted; they are precise to the second decimal as their errors mostly reflect the fluctuations of the local red+white noise levels (see Table \ref{paramrn}) which are at 0.10--0.15\,mmag for WR\,7 and WR\,66, 0.01--0.04\,mmag for WR79b, 0.6--1.0\,mmag for WR\,134, and 0.03--0.10\,mmag for WR135 (the largest value corresponds to the lowest frequency). } 
\end{table*}

The computed time-frequency diagrams (Fig. \ref{tfdiag2}) show that these signals are stable in frequency over the duration of the \te\ observations, although their amplitudes may somewhat change. The only exception seems to be WR\,134, for which the low-frequency periodogram shows continuous changes. The presence of strong red noise however renders more difficult to assess the stability of the low-frequency signal found for this star. In this context, it is important to note that \citet{mcc94}, \citet{gos96} and \citet{mor99} found a periodicity of 0.44\,d$^{-1}$ from line profile variability in spectroscopic datasets of WR\,134, a value close to that observed in \te\ data: this signal may thus be long-lived. 

The peaks appear complex for the one-sector cases (WR\,7, 66, and 79b), i.e. peaks are flanked by subpeaks. Such aliases probably arise from the sampling, which notably creates sidelobes at $\nu_{\rm peak}\pm0.06\,{\rm d}^{-1}$ (Fig. \ref{freq}) - asserting the presence of actual subpeaks will require a better frequency resolution, i.e. longer lightcurves. What can however already be tested are the relations between frequencies. The main frequencies of WR\,7 and those of WR\,134 and 135 clearly are harmonics, while additional combinations of signals possibly exist in WR\,7 (a faint signal near 1.046 is close to the difference between 6.660 and 5.628\,d$^{-1}$) and WR\,66 (12.996$\sim$6.076+6.928\,d$^{-1}$). Apart from these, there seems to be no direct relationship (including equal spacing) between signals: WR\,7, 66, and 79b thus truly display multiple periodicities. 

Several searches for high-frequency signals were performed in the past for WR stars, notably in the context of searches for compact companions \citep[e.g.][]{mar94}. However, there were few cases of reported and confirmed periodicities. Focusing on our sample, the following detections were published. \citet{ble92} claimed a detection of pulsations with a 627\,s period and a 5\,mmag peak-to-peak amplitude in WR\,40. However, at the corresponding frequency of 138\,d$^{-1}$, the \te\ high-cadence data (sector 10) show no sign of such a signal, confirming previous negative reports \citep{gos94,mar94,nat94,sch94}, including that by the discovery team itself \citep{bra96}. Other low-frequency detections for that star \citep[e.g.][]{ant95} certainly correspond to the red noise stochastic variability. Another photometric campaign detected a 6.828\,d$^{-1}$ signal in WR\,66 \citep{ant95} but with strong daily aliasing. It was subsequently confirmed in an independent, less aliased dataset \citep[largest peak at 5.815\,d$^{-1}$][]{rau96}. These values are close (but not identical) to the frequency of the largest \te\ peak, or its daily alias. To assess the compatibility between datasets, we have simulated a signal composed of the three main frequencies detected by \te\ but sampled as in \citet{rau96}. The resulting periodogram appears similar to the one reported in the literature, even if the passbands are different\footnote{\citet{rau96} used Str\"omgren $b$ photometry, which is affected by the presence of a strong He\,{\sc ii}\,4686\AA\ line, whereas the \te\ passband is larger and redder.}: the old and new datasets therefore appear fully compatible.  Note also that the presence of several frequencies casts further doubt on the hypothesis of the literature signal being the orbital period of a compact companion \citep{ant95}. Finally, a 25\,min signal with 2.6\,mmag amplitude was reported by \citet{bra96} for WR\,78 but considered as a transient event as it was observed only during one night. Because WR\,78 was not observed with 2\,min cadence, we cannot check the presence of such a frequency in the \te\ data.

Except for WR\,134, the detected frequencies are high, from 3\,d$^{-1}$ up to 14\,d$^{-1}$. This cannot be easily reconciled with an orbital period (the companion would travel inside the WR star) nor a rotation rate (it would be above break-up velocity). Therefore, the most probable culprits are pulsations. WR\,66 very probably possesses a close neighbour (see Sect. 2.1), which could cast doubt on the identification of the pulsating star, but that is not the case of the other stars (including WR\,123, see \citealt{lef05}). Pulsations here are thus considered to originate from the WR star itself, and such a possibility has actually been considered before. Pushed by the observational considerations of \citet{vre85}, the first pulsation models for WRs were elaborated more than thirty years ago, with predicted periods of about one hour \citep{mae85,scu86}. Subsequent work by \citet{gla99} focused on strange modes in small and hot helium stars (often taken as behaving similarly to WRs) and the predicted pulsations had periods of several minutes with amplitudes of several mmag. After the discovery of a 2.45\,d$^{-1}$ signal in WR\,123 \citep{lef05}, i.e. at a much lower frequency than expected, the models were revisited. Strange modes were extended to stars with larger radii \citep{dor06,gla08} while predictions of g-modes excited by the $\kappa$ mechanism were made for WRs by \citet{tow06}. In both cases, the predicted periods could be made compatible with the 10\,hr signal observed in WR\,123. The signals we observe have frequencies in the same range, even for the 2\,min cadence data, which allows to probe very high frequencies, no coherent signal is detected above 14\,d$^{-1}$: WRs thus seem to pulsate only at moderately high frequencies. Furthermore, \citet{tow06} expected pulsations with higher frequencies in early WN (2--8\,d$^{-1}$) than in late WN (1--2\,d$^{-1}$). For our pulsators, only WR\,7 has an early WN type and its frequency values are not particularly higher: its main signals are in the same frequency range as those of the late-type WR\,66. Moreover, the highest frequencies are all found in late WN stars. However, \citet{tow06} models were made for ``general/typical'' stars, not specific ones. In addition, since the WR light mostly comes from inside the wind, the filtering effect by the wind on a signal from the underlying hydrostatic surface needs to be assessed in detail. New, dedicated models will be needed for a more in-depth comparison between observations and predictions.

\section{Conclusion}
In this paper, we aim at characterizing the high-frequency variability of evolved massive stars, either WR stars or LBVs (including LBV candidates). To avoid confusion, we selected only stars not known to be multiple and without bright ($\Delta G<2.5\,mag$) and close (within 1\arcmin ) neighbours in the {\it Gaia}-DR2 catalog. Of these, 26 WRs and 8 (c)LBVs had available \te\ photometry, with 2\,min cadence in 7 cases and 30\,min cadence otherwise.

All lightcurves display low-frequency stochastic variability in addition to overall white noise. The parameters of this red noise (level, slope and position of the transition towards white noise) cover a similar range as found for OB-stars by \citet{bow20}. The white noise level appears larger than in OB-stars, although when focusing on massive stars with similar luminosities and temperatures, the difference is reduced. Few significant correlations are found, however: as WC stars brighten, the noise levels decrease while the transition red$\rightarrow$white noise appears somewhat slower for hotter WRs. 

No coherent, isolated signal is found for (c)LBVs, but such signals are detected in five WRs: WR\,7, 66, 79b, 134, and 135. WR\,134 shows a signal and its first harmonic, WR\,135 displays one frequency and its first two harmonics, while the last three stars appear multiperiodic. One frequency in WR\,66 and one in WR\,134 were reported previously and are thus confirmed by \te\ data. Except for WR\,134, the detected signals appear at high frequencies (3--14\,d$^{-1}$): such values exclude orbital and rotational modulations, rather favoring a pulsational origin. Our results thus add WR\,7, 79b, and 135 to the list of known Galactic high-frequency pulsators which contained up to now WR\,66 and 123. Besides, there is no clear trend with WR subtype, unlike what is sometimes predicted by models. Dedicated modelling is now needed to understand the stellar properties at the hydrostatic surface as well as the exact role of the overlying wind, such as filtering effect and/or additional source of variability.

\section*{Acknowledgements}
The authors acknowledge support from the Fonds National de la Recherche Scientifique (Belgium), the European Space Agency (ESA) and the Belgian Federal Science Policy Office (BELSPO) in the framework of the PRODEX Programme (contracts linked to XMM-Newton and Gaia). They also thank the TESS helpdesk (in particular R.A. Hounsell) for discussion and advices, and D.M. Bowman for discussion. ADS and CDS were used for preparing this document. 

\section*{Data availability}
The \te\ data used in this article are available in the MAST archives.

\bsp	% typesetting comment
\label{lastpage}

\begin{thebibliography}{99}
%\bibitem[\protect\citeauthoryear{Antokhin et al.}{1995}]{ant95} Antokhin, I.~I., Bertrand, J.-F., Lamontagne, R., et al.\ 1995, Wolf-Rayet Stars: Binaries; Colliding Winds; Evolution, 163, 62
\bibitem[\protect\citeauthoryear{Antokhin et al.}{1995}]{ant95} Antokhin, I., Bertrand, J.-F., Lamontagne, R., et al.\ 1995, \aj, 109, 817
\bibitem[\protect\citeauthoryear{Balona \& Ozuyar}{2020}]{bal20} Balona, L.~A., \& Ozuyar, D.\ 2020, \mnras, 493, 2528
\bibitem[\protect\citeauthoryear{Baran et al.}{2015}]{bar15} Baran, A.~S., Koen, C., \& Pokrzywka, B.\ 2015, \mnras, 448, L16
\bibitem[\protect\citeauthoryear{Blecha et al.}{1992}]{ble92} Blecha, A., Schaller, G., \& Maeder, A.\ 1992, \nat, 360, 320
\bibitem[\protect\citeauthoryear{Blomme et al.}{2011}]{blo11} Blomme, R., Mahy, L., Catala, C., et al.\ 2011, \aap, 533, A4
\bibitem[\protect\citeauthoryear{Briquet et al.}{2011}]{bri11} Briquet, M., Aerts, C., Baglin, A., et al.\ 2011, \aap, 527, A112
\bibitem[\protect\citeauthoryear{Bohannan \& Crowther}{1999}]{boh99} Bohannan, B. \& Crowther, P.~A.\ 1999, \apj, 511, 374
\bibitem[\protect\citeauthoryear{Bowman et al.}{2019a}]{bowcorot} Bowman, D.~M., Aerts, C., Johnston, C., et al.\ 2019a, \aap, 621, A135
\bibitem[\protect\citeauthoryear{Bowman et al.}{2019b}]{bow19} Bowman, D.~M., Burssens, S., Pedersen, M.~G., et al.\ 2019b, Nature Astronomy, 3, 760
\bibitem[\protect\citeauthoryear{Bowman et al.}{2020}]{bow20}  Bowman, D.M., Burssens, S., Sim\'{o}n-D\'{\i}az, S., et al.\ 2020, \aap, 640, A36
\bibitem[\protect\citeauthoryear{Bratschi \& Blecha}{1996}]{bra96} Bratschi, P. \& Blecha, A.\ 1996, \aap, 313, 537
\bibitem[\protect\citeauthoryear{Chen{\'e} et al.}{2011}]{che11} Chen{\'e}, A.-N., Moffat, A.~F.~J., Cameron, C., et al.\ 2011, \apj, 735, 34
\bibitem[\protect\citeauthoryear{David-Uraz et al.}{2012}]{dav12} David-Uraz, A., Moffat, A.~F.~J., Chen{\'e}, A.-N., et al.\ 2012, \mnras, 426, 1720
\bibitem[\protect\citeauthoryear{Deeming}{1975}]{dee75} Deeming, T.~J.\ 1975, \apss, 36, 137. doi:10.1007/BF00681947
\bibitem[\protect\citeauthoryear{Degroote et al.}{2010}]{deg10} Degroote, P., Briquet, M., Auvergne, M., et al.\ 2010, \aap, 519, A38
\bibitem[\protect\citeauthoryear{Dorfi et al.}{2006}]{dor06} Dorfi, E.~A., Gautschy, A., \& Saio, H.\ 2006, \aap, 453, L35
\bibitem[\protect\citeauthoryear{Gaia Collaboration et al.}{2016}]{gaia16} Gaia Collaboration, Prusti, T., de Bruijne, J.~H.~J., et al.\ 2016, \aap, 595, A1
\bibitem[\protect\citeauthoryear{Gaia Collaboration et al.}{2018}]{gaia18} Gaia Collaboration, Brown, A.~G.~A., Vallenari, A., et al.\ 2018, \aap, 616, A1
\bibitem[\protect\citeauthoryear{Glatzel et al.}{1999}]{gla99} Glatzel, W., Kiriakidis, M., Chernigovskij, S., et al.\ 1999, \mnras, 303, 116
\bibitem[\protect\citeauthoryear{Glatzel}{2008}]{gla08} Glatzel, W.\ 2008, in proceedings of ``Hydrogen-Deficient Stars'', ASP Conference Series, Vol. 391, p.307
\bibitem[\protect\citeauthoryear{Godart et al.}{2017}]{god17} Godart, M., Sim{\'o}n-D{\'\i}az, S., Herrero, A., et al.\ 2017, \aap, 597, A23
\bibitem[\protect\citeauthoryear{Gosset \& Vreux}{1996}]{gos96} Gosset, E. \& Vreux, J.~M.\ 1996, Liege International Astrophysical Colloquia, 33, 231
\bibitem[\protect\citeauthoryear{Gosset et al.}{1990}]{gos90} Gosset, E., Vreux, J.-M., Manfroid, J., et al.\ 1990, \aaps, 84, 377
\bibitem[\protect\citeauthoryear{Gosset et al.}{1994}]{gos94} Gosset, E., Rauw, G., Manfroid, J., et al.\ 1994, NATO Advanced Science Institutes (ASI) Series C, 436, 101
\bibitem[\protect\citeauthoryear{Gosset et al.}{2001}]{gos01} Gosset, E., Royer, P., Rauw, G., et al.\ 2001, \mnras, 327, 435
\bibitem[\protect\citeauthoryear{Gr{\"a}fener et al.}{2017}]{gra17} Gr{\"a}fener, G., Owocki, S.~P., Grassitelli, L., et al.\ 2017, \aap, 608, A34
\bibitem[\protect\citeauthoryear{Gvaramadze et al.}{2009}]{gva09} Gvaramadze, V.~V., Fabrika, S., Hamann, W.-R., et al.\ 2009, \mnras, 400, 524
%\bibitem[\protect\citeauthoryear{Hamann et al.}{2006}]{ham06} Hamann, W.-R., Gr{\"a}fener, G., \& Liermann, A.\ 2006, \aap, 457, 1015
\bibitem[\protect\citeauthoryear{Hamann et al.}{2019}]{ham19} Hamann, W.-R., Gr{\"a}fener, G., Liermann, A., et al.\ 2019, \aap, 625, A57
\bibitem[\protect\citeauthoryear{Harvey}{1985}]{har85} Harvey, J.\ 1985, Future Missions in Solar, Heliospheric \& Space Plasma Physics, 235, 199
\bibitem[\protect\citeauthoryear{Heck et al.}{1985}]{hmm} Heck, A., Manfroid, J., \& Mersch, G.\ 1985, \aaps, 59, 63 
\bibitem[\protect\citeauthoryear{Kallinger et al.}{2014}]{kal14} Kallinger, T., De Ridder, J., Hekker, S., et al.\ 2014, \aap, 570, A41. doi:10.1051/0004-6361/201424313
\bibitem[\protect\citeauthoryear{Lef{\`e}vre et al.}{2005}]{lef05} Lef{\`e}vre, L., Marchenko, S.~V., Moffat, A.~F.~J., et al.\ 2005, \apjl, 634, L109
\bibitem[\protect\citeauthoryear{Lenoir-Craig et al.}{2020}]{len20} Lenoir-Craig, G., St-Louis, N., Moffat, A.~F.~J., et al.\ 2020, Proceedings of the conference Stars and their Variability Observed from Space, 191
\bibitem[\protect\citeauthoryear{Lindegren}{2016}]{lin16}Lindegren, L.\ 2016, Re-normalising the astrometric chi-square in Gaia DR2, technical note, GAIA-C3-TN-LU-LL-124-01, available from  http://www.rssd.esa.int/doc\_fetch.php?id=3757412
\bibitem[\protect\citeauthoryear{Lucy \& Abbott}{1993}]{luc93} Lucy, L.~B. \& Abbott, D.~C.\ 1993, \apj, 405, 738
\bibitem[\protect\citeauthoryear{McCandliss et al.}{1994}]{mcc94} McCandliss, S.~R., Bohannan, B., Robert, C., et al.\ 1994, \apss, 221, 155
\bibitem[\protect\citeauthoryear{Maeder}{1985}]{mae85} Maeder, A.\ 1985, \aap, 147, 300
\bibitem[\protect\citeauthoryear{Mahy et al.}{2011}]{mah11} Mahy, L., Gosset, E., Baudin, F., et al.\ 2011, \aap, 525, A101
\bibitem[\protect\citeauthoryear{Marchenko et al.}{1994}]{mar94} Marchenko, S.~V., Antokhin, I.~I., Bertrand, J.-F., et al.\ 1994, \aj, 108, 678
\bibitem[\protect\citeauthoryear{Martinez et al.}{1994}]{nat94} Martinez, P., Kurtz, D., Ashley, R., et al.\ 1994, \nat, 367, 601
\bibitem[\protect\citeauthoryear{Moffat et al.}{1988}]{mof88} Moffat, A.~F.~J., Drissen, L., Lamontagne, R., et al.\ 1988, \apj, 334, 1038
\bibitem[\protect\citeauthoryear{Moffat et al.}{2008}]{mof08} Moffat, A.~F.~J., Marchenko, S.~V., Lef{\`e}vre, L., et al.\ 2008, Mass Loss from Stars and the Evolution of Stellar Clusters, 388, 29
%\bibitem[\protect\citeauthoryear{Moffat et al.}{2018}]{mof18} Moffat, A.~F.~J., St-Louis, N., Carlos-Leblanc, D., et al.\ 2018, 3rd BRITE Science Conference, 8, 37
\bibitem[\protect\citeauthoryear{Morel et al.}{1999}]{mor99} Morel, T., Marchenko, S.~V., Eenens, P.~R.~J., et al.\ 1999, \apj, 518, 428
\bibitem[\protect\citeauthoryear{Naz{\'e} et al.}{2012}]{naz12} Naz{\'e}, Y., Rauw, G., \& Hutsem{\'e}kers, D.\ 2012, \aap, 538, A47
\bibitem[\protect\citeauthoryear{Rauw et al.}{1996}]{rau96} Rauw, G., Gosset, E., Manfroid, J., et al.\ 1996, \aap, 306, 783
\bibitem[\protect\citeauthoryear{Rauw et al.}{2019}]{rau19} Rauw, G., Pigulski, A., Naz{\'e}, Y., et al.\ 2019, \aap, 621, A15
\bibitem[\protect\citeauthoryear{Ramiaramanantsoa et al.}{2018}]{Tahina}  Ramiaramanantsoa, T., Moffat, A.F.J., Harmon, R., et al.\ 2018, \mnras, 473, 5532
\bibitem[\protect\citeauthoryear{Ramiaramanantsoa et al.}{2019}]{ram19} Ramiaramanantsoa, T., Ignace, R., Moffat, A.~F.~J., et al.\ 2019, \mnras, 490, 5921
\bibitem[\protect\citeauthoryear{Richardson et al.}{2011}]{ric11} Richardson, N.~D., Gies, D.~R., \& Williams, S.~J.\ 2011, \aj, 142, 201
\bibitem[\protect\citeauthoryear{Richardson et al.}{2016}]{ric16} Richardson, N.~D., Moffat, A.~F.~J., Maltais-Tariant, R., et al.\ 2016, \mnras, 455, 244
\bibitem[\protect\citeauthoryear{Ricker et al.}{2015}]{ric15} Ricker, G.~R., Winn, J.~N., Vanderspek, R., et al.\ 2015, Journal of Astronomical Telescopes, Instruments, and Systems, 1, 014003
\bibitem[\protect\citeauthoryear{Rogers et al.}{2013}]{Rog13} Rogers, T.M., Lin, D.N.C., McElwaine, J.N., \& Lau, H.B.B.\ 2013, \apj, 772, 21
%\bibitem[\protect\citeauthoryear{Sander et al.}{2012}]{san12} Sander, A., Hamann, W.-R., \& Todt, H.\ 2012, \aap, 540, A144
\bibitem[\protect\citeauthoryear{Sander et al.}{2019}]{san19} Sander, A.~A.~C., Hamann, W.-R., Todt, H., et al.\ 2019, \aap, 621, A92
\bibitem[\protect\citeauthoryear{Scargle}{1982}]{sca82} Scargle, J.~D.\ 1982, \apj, 263, 835. doi:10.1086/160554
\bibitem[\protect\citeauthoryear{Schmutz \& Koenigsberger}{2019}]{sch19} Schmutz, W. \& Koenigsberger, G.\ 2019, \aap, 624, L3
\bibitem[\protect\citeauthoryear{Schneider et al.}{1994}]{sch94} Schneider, H., Kiriakidis, M., Weiss, W.~W., et al.\ 1994, Pulsation; Rotation; and Mass Loss in Early-Type Stars, 162, 53
\bibitem[\protect\citeauthoryear{Schneider et al.}{1997}]{sch97} Schneider, H., Glatzel, W., \& Fricke, K.~J.\ 1997, Luminous Blue Variables: Massive Stars in Transition, 120, 206
\bibitem[\protect\citeauthoryear{Scuflaire \& Noels}{1986}]{scu86} Scuflaire, R. \& Noels, A.\ 1986, \aap, 169, 185
\bibitem[\protect\citeauthoryear{Stanishev et al.}{2002}]{Stanishev} Stanishev, V., Kraicheva, Z., Boffin, H.M.J., \& Genkov, V.\ 2002, \aap, 394, 625
\bibitem[\protect\citeauthoryear{Sterken \& Breysacher}{1997}]{ste97} Sterken, C. \& Breysacher, J.\ 1997, \aap, 328, 269
\bibitem[\protect\citeauthoryear{Townsend \& Owocki}{2005}]{tow05} Townsend, R.~H.~D. \& Owocki, S.~P.\ 2005, \mnras, 357, 251
\bibitem[\protect\citeauthoryear{Townsend \& MacDonald}{2006}]{tow06} Townsend, R.~H.~D. \& MacDonald, J.\ 2006, \mnras, 368, L57
\bibitem[\protect\citeauthoryear{van der Hucht}{2001}]{vdh01} van der Hucht, K.~A.\ 2001, \nar, 45, 135
\bibitem[\protect\citeauthoryear{Vreux}{1985}]{vre85} Vreux, J.-M.\ 1985, \pasp, 97, 274
\bibitem[\protect\citeauthoryear{Zechmeister \& K\"urster}{2009}]{zec09} Zechmeister, M. \& K\"urster, M.\ 2009, \aap, 496, 577
\end{thebibliography}
\end{document}